\documentclass[trackchanges,twocolumn]{aastex7}

\usepackage{csvsimple}
\usepackage{booktabs}
\usepackage{longtable}
\usepackage{multirow}
\usepackage{amsmath}

\usepackage[euler]{textgreek}
\accepted{for Publication in The Astrophysical Journal Letters}

\begin{document}


\title{Giant impacts preferentially remove low-mass atmospheres: a generalised scaling law for impact-driven atmospheric loss}

\author[orcid=0000-0002-2595-1393]{Matthew J. Roche}
\affiliation{School of Earth Sciences, Wills Memorial Building, University of Bristol, Bristol, BS8 1RJ, UK}
\email{matthew.roche@bristol.ac.uk} 

\author[0000-0001-5365-9616]{Simon J. Lock}
\affiliation{School of Earth Sciences, Wills Memorial Building, University of Bristol, Bristol, BS8 1RJ, UK}
\email[]{s.lock@bristol.ac.uk}

\author[0000-0001-5065-4625]{Philip J. Carter}
\affiliation{School of Physics, HH Wills Physics Laboratory, University of Bristol, Bristol, BS8 1TL, UK}
\email[]{p.carter@bristol.ac.uk}

\author[0000-0003-4813-7922]{Zo\"{e} M. Leinhardt}
\affiliation{School of Physics, HH Wills Physics Laboratory, University of Bristol, Bristol, BS8 1TL, UK}
\email[]{Zoe.Leinhardt@bristol.ac.uk}

\correspondingauthor{Matthew J. Roche}
\email{matthew.roche@bristol.ac.uk}

\begin{abstract} 
\noindent Impact-driven atmospheric loss plays a key role in shaping the diversity of planetary atmospheres.
However, existing prescriptions for calculating loss from a given impact neglect the critical role of atmospheric mass on loss, and thus often severely miscalculate the extent of loss.
Using 3D smoothed particle hydrodynamics simulations, we quantify atmospheric loss during giant impacts onto planets with 1--20\% mass fraction atmospheres and show that less massive (lower-pressure) atmospheres are significantly easier to remove.
We derive a new scaling law for atmospheric loss that includes the effects of atmospheric mass and is applicable to collisions between volatile-poor planetary embryos through to super-Earths and sub-Neptunes. 
Applying our scaling law to $N$-body simulations of solar system formation, we show that including the coupling between atmospheric mass and loss efficiency results in significantly more atmosphere removal. 
Substantial atmospheric loss events are common for smaller planets ($\lesssim$0.6 M$_{\oplus}$), whilst even roughly Earth mass bodies lose 30--100\% of their atmospheres through multiple impacts.
We also show that atmospheric loss during the Moon-forming impact strongly depends on both the pre-impact atmospheric mass and impact style. 
For a representative $\sim$100 bar pre-impact atmosphere, Earth could have lost between 20--80\% of its atmosphere.
In any system that undergoes a phase of giant impacts, the atmospheres of its host planets will be unavoidably shaped by the stochasticity of impacts.
Our new scaling law can be readily incorporated into models of planet formation and system-wide evolution, offering new insights into the origins of planetary diversity.

\end{abstract}



\section{Introduction} \label{sec:intro}

In many systems, terrestrial planet formation concludes with sequences of giant impacts \citep[e.g.,][]{agnor_character_1999,quintana2016frequency} -- violent collisions between protoplanets.
Giant impacts are sufficiently energetic to melt and vaporise large portions of the colliding mass \citep[][]{davies_silicate_2020,carter2020energy,caracas2023no}, and can eject significant amounts of atmosphere and ocean \citep[][]{genda2005enhanced,kegerreis_atmospheric_2020,denman2022atmosphere,lock2024atmospheric,roche_atmospheric_2025}.
Impact-driven atmospheric loss has been invoked to partly explain (a) the exoplanet `radius valley' \citep[e.g.,][]{matsumoto_size_2021,chance2022signatures,shibata_tale_2025}, (b) density contrasts between similarly-sized planets in the same system \citep{liu2015giant,inamdar2016stealing,bonomo2019giant}, and (c) Earth's depletion in highly volatile elements relative to its chondritic building blocks \citep{tucker2014evidence,sakuraba_numerous_2021,lock2024atmospheric}.
Precisely quantifying the efficiency of atmospheric loss across varied impact conditions is therefore crucial to constraining the role of giant impacts in shaping the volatile element budgets of planets, and ultimately whether planets can retain atmospheres and oceans conducive to their habitability.

Cutting-edge 3D smoothed particle hydrodynamics (SPH) simulations can now resolve increasingly thin atmospheres \citep[e.g.,][]{kegerreis_atmospheric_2020} and capture atmospheric loss via all direct impact-driven mechanisms -- air shocks, vaporised ejecta plumes, and shock kick \citep[][]{roche_atmospheric_2025}.
These advances have enabled the development of scaling laws that predict atmospheric loss for a given impact scenario \citep[][]{kegerreis_atmospheric_2020,denman2022atmosphere,roche_atmospheric_2025}, forming key components of planet-/system-wide evolution models \citep[e.g,][]{matsumoto_size_2021,gu2024composition}.
However, impact-driven atmospheric loss is highly dependent on atmospheric properties, namely composition, surface pressure, and temperature, as well as the presence of surface oceans \citep[][]{lock2024atmospheric,genda2005enhanced}.
In particular, lower-pressure (less massive) atmospheres are thought to be substantially more susceptible to loss \citep{lock2024atmospheric}.
Yet existing scaling laws \citep{kegerreis_atmospheric_2020,denman2022atmosphere,roche_atmospheric_2025} were derived from simulation suites that included impacts with only one target atmosphere mass fraction and do not account for the dependency of loss efficiency on atmosphere mass. 
As a result, existing scaling laws can severely miscalculate atmospheric loss, limiting their applicability in planetary evolution models.

Here, we present the results of 300 new 3D SPH simulations onto roughly Earth-mass planets with 20\%, 10\%, and 1\% mass fraction H$_2$--He atmospheres spanning a range of impact angles, velocities, and impactor-to-system mass ratios.
We combine these results with the suite of impact simulations from \citet{roche_atmospheric_2025} that simulated 5\% mass fraction H$_2$--He atmospheres and additionally explored the dependency of loss on target planet masses.
From this combined dataset, we derive an improved scaling law for atmospheric loss that includes the dependence on atmospheric mass. 
The resulting prescription is widely applicable for use on impacts between small, volatile-poor planetary embryos through to super-Earths and sub-Neptunes with thick volatile envelopes.
An implementation of our scaling laws in \texttt{python} and plots to reproduce the figures in this manuscript are available on GitHub.

\section{Methods} \label{sec:methods}

\subsection{Initialising the colliding bodies} \label{subsec:init_cond}
For the new simulations, we considered impacts onto non-rotating target planets with refractory masses ($M_{\rm t}^{\rm r}$, i.e., neglecting atmosphere) of 1.0 Earth mass (M$_{\oplus}$) and three different atmosphere mass fractions, $f_{\rm atm}$, with five impactor-to-total system mass ratios (Section \ref{subsec:impact_sims}).
No impactors had atmospheres. 

We used the \texttt{WoMa} (version 1.2.0) planetary structure code \citep{ruiz2021effect} to generate 1D interior thermal profiles for pre-impact planets. 
All planets were differentiated into a metal core and silicate mantle with a core mass fraction of $\sim$0.3, modelled using M-ANEOS pure iron \citep{sarah_t_stewart_2020_3866507} and forsterite \citep{stewart_2019_3478631,10.1063/12.0000946} equations of state, respectively. 
Cores and mantles were modelled as isentropic, with a mantle specific entropy of $\sim$2.6 kJ kg$^{-1}$K$^{-1}$ (corresponding to a 1~bar potential temperature of $\sim$1700 K) and a core specific entropy of $\sim$1.8 kJ kg$^{-1}$K$^{-1}$, (resulting in core--mantle boundary temperatures of 3000--4000~K, depending on planet mass).
Target planets were initialised with 1\%, 10\%, and 20\% mass fraction H$_2$--He atmospheres at constant internal energy using the EoS from \citet{hubbard1980structure}, each with a fixed basal temperature of 500 K, which resulted in mantle surface pressures of 0.71, 3.10, and 4.62 GPa, and total target planet masses ($M_{\rm t}^{\rm tot}$) of 1.01, 1.11, and 1.25 M$_{\oplus}$, respectively (Table \ref{table:initial_conditions}). 

Approximately equal mass SPH particles were then arranged to construct 3D planets with the calculated 1D thermal profiles using the stretched equal-area method \citep[][\texttt{SEAGen} version 1.4.2]{kegerreis2019planetary}.
Planet initialisation was concluded by running settling simulation for each isolated body. 
The particles' specific entropies (or specific internal energies for atmosphere particles since the \citet{hubbard1980structure} EoS does not natively include entropy) were fixed for the first 10 hrs and all parameters allowed to evolve freely for another 10 hrs.

\subsection{Impact simulations} \label{subsec:impact_sims} 
We considered impact scenarios for every combination of: (i) impact parameter, $b \in \{0.0, 0.3, 0.5, 0.7, 0.9\}$; (ii) velocity, $v_{\rm c} \in \{1.0, 1.5, 2.0, 3.0\}v_{\rm esc}$, where $v_{\rm esc}$ is the mutual escape velocity between the colliding bodies at the point of contact between the two mantles:

\begin{equation}
    v_{\rm esc} = \sqrt{2G \left(\frac{M_{\rm t}^{\rm tot} + M_{\rm i}^{\rm r}}{R_{\rm t}^{\rm r} + R_{\rm i}^{\rm r}}\right)} \;,
\end{equation}

\noindent with $G$ being the gravitational constant, and $R_{\rm t}^{\rm r}$ and $R_{\rm i}^{\rm r}$ the refractory target and impactor radii; (iii) impactor-to-total (refractory) system mass ratio, $\gamma \in \{0.1, 0.2, 0.3, 0.4, 0.5\}$:

\begin{equation}
    \gamma = \frac{M_{\rm i}^{\rm r}}{M_{\rm i}^{\rm r} + M_{\rm t}^{\rm r}} \;;
\end{equation}

\noindent and (iv) target atmosphere mass fraction, $f_{\rm atm} \in \{0.01, 0.1, 0.2\}$:

\begin{equation}\label{eq:f_atm}
    f_{\rm atm} = \frac{M_{\rm t}^{\rm tot} - M_{\rm t}^{\rm r}}{M_{\rm t}^{\rm tot}} \; ,
\end{equation}

\noindent (Table \ref{table:initial_conditions}).
The impact parameter and velocity were defined using the point at which the mantles of the colliding bodies first touch as the point of impact, allowing for comparison between scenarios where the targets had different atmosphere mass fractions and thus different total radii. 
The positions and velocities of the impactor and target were set such that collision of the mantles occured $\sim$1~hr after the start of the simulation  using the inbuilt solver in \texttt{WoMa}. 
\citet{roche_atmospheric_2025} found that the choice of pre-collision approach time made little difference to the total amount of atmospheric loss, at least when considering fluid planets. 

We ran a total of 300 new impact simulations with $\sim$10$^6$ SPH particles per target mass for $f_{\rm atm} \in \{0.1, 0.2\}$ and $\sim$10$^{6.5}$ for $f_{\rm atm} = 0.01$ to ensure that the atmosphere of each target was modelled using at least 5 particle shells, as \citet{roche_atmospheric_2025} showed atmospheric loss converged for simulations with $\geq$5 particle shells in the atmosphere and typically with $\geq$10$^{5.5}$ particles per total target planet mass. 
The number of particles in each impactor was scaled to give approximately the same mass particles in each colliding body.
Although prior works have suggested much higher resolutions are required for convergence of impact outcomes \citep[e.g.,][]{kegerreis2019planetary}, it is known that the choice of equation of state can significantly effect numerical convergence \citep{meier2021eos} -- reasoning suggested by \citet{roche_atmospheric_2025} for discrepancies between the number of particles required to accurately resolve atmosphere layers compared to previous work \citep[e.g.,][]{kegerreis2019planetary,kegerreis_atmospheric_2020}.
In the case of atmospheric loss however, convergence between final atmospheric loss fractions at resolutions equal to or greater than those considered here are typically on the order a percent \citep{kegerreis2019planetary,kegerreis_atmospheric_2020}.
Such small differences are to be expected given the chaotic nature of giant impact simulations and would not affect the overall conclusions of this work.
All SPH simulations were run with the \texttt{SWIFT} (version 0.9.0) code using the \texttt{subtask\_speedup} branch \citep{kegerreis2019planetary,schaller_swift_2024}. 
The version of the code used is archived on Zenodo \citep{schaller_mjr230600swiftsim_2024}.
All simulations were run using a cubic spline kernel with the \citet{balsara1995neumann} switch for artificial viscosity and used a maximum smoothing length, $h_{\rm max}$, of 0.1 R$_{\oplus}$, which set a minimum density of 1.74--11.9 kg m$^{-3}$.
Each simulation used a cubic boxsize of 1000 R$_{\oplus}$ and was run for a total of 10 hrs, since \citet{roche_atmospheric_2025} showed convergence of atmospheric loss after $\sim$6 hrs. 

In addition to our new simulations, we also make use of simulations from \citet{roche_atmospheric_2025} that calculated loss from target planets with $f_{\rm atm}=0.05$ (i.e., 5\% mass fraction) for the same impact parameters considered here, but over a number of total target planet masses, $M_{\rm t}^{\rm tot}$ $\in$ \{0.37, 1.05, 2.1, 5.25\}~M$_{\oplus}$. 
The impacts were initialised, run, and analysed following the same procedure as our new simulations. 

\subsection{Calculating combined, near-, and far-field atmospheric loss} \label{subsec:calc_loss}
We calculated atmospheric loss following the same approach as \citet{roche_atmospheric_2025}. 
We tracked the change in atmosphere mass gravitationally bound to the largest remnant at each time step, normalised to the initial atmospheric mass.
Also, we separated the pre-impact target planet's atmosphere into near- and far-field components using the tangent surface method  \citep{roche_atmospheric_2025}, allowing us to de-convolve the atmospheric loss into near- (via ejecta plumes and air shocks) and far-field (via shock-kick) components, helping us understand the controls on loss and produce accurate scaling laws.

\section{Results} \label{sec:results}

\begin{figure*}
\begin{center}
    \includegraphics[width=\textwidth]{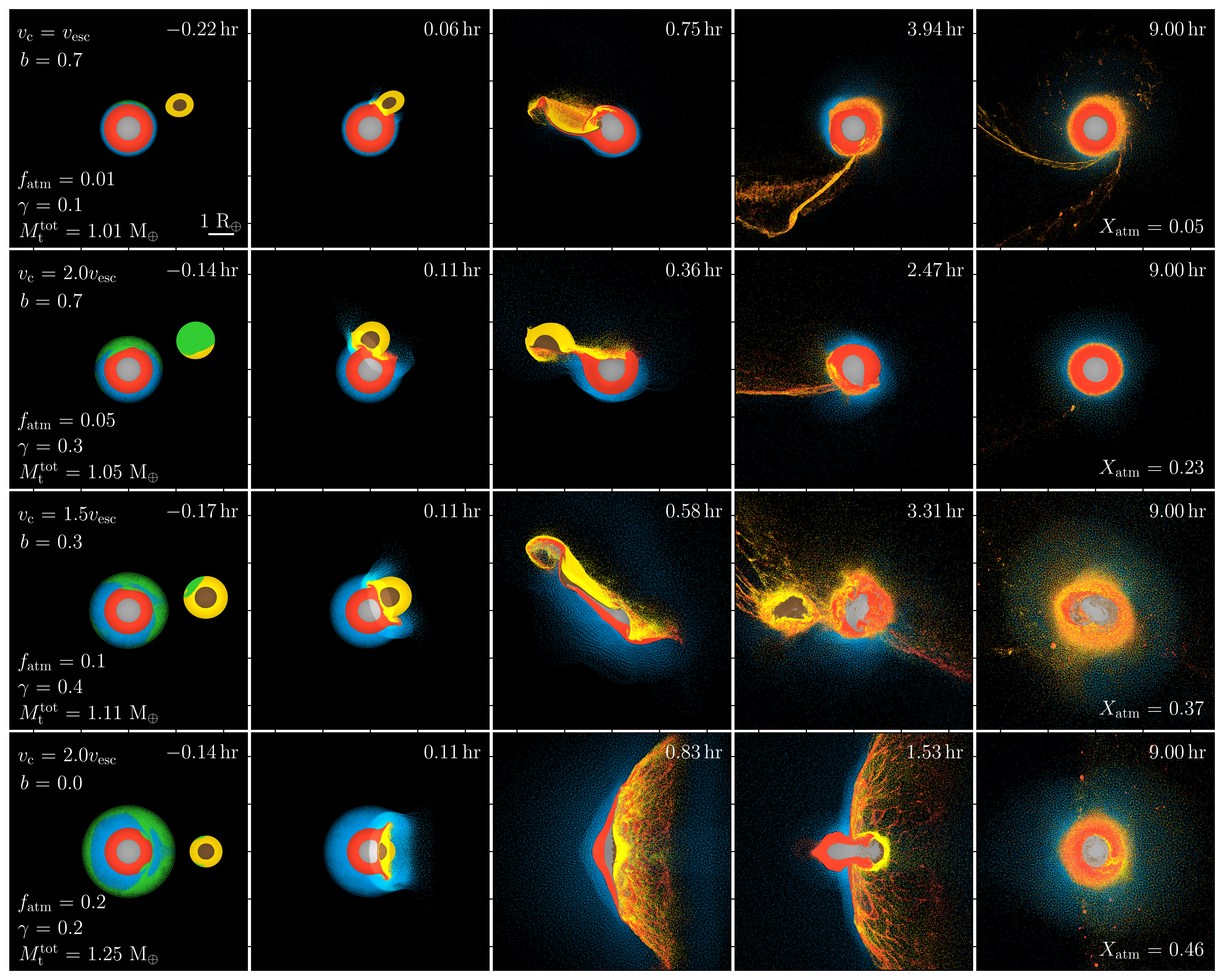}
    \caption{Illustrative time snapshots (columns) of four separate SPH simulations with different atmosphere mass fractions (rows), where the particle resolution is $\sim$10$^6$ for all except $f_{\rm atm}=0.01$ where it is $\sim$10$^{6.5}$. Particles in the lower hemispheres of the colliding bodies are shown and are coloured by their material types: target core (grey), mantle (orange), and atmosphere (blue), and impactor core (brown) and mantle (yellow). Green particles in the first column indicate those that ultimately become unbound (`lost') from the system. Impact parameters are given in the first column with the simulation time in each panel referenced to $t=0$ hrs being the point of first contact of the colliding bodies' mantles. Atmospheric loss fractions are given in the final column. An animation of this figure is available in the online article. The 12 s animation spans simulation times from --0.97 to 9.0 hrs.}
    \label{fig:impacts}
\end{center}
\end{figure*}

Figure \ref{fig:impacts} shows time snapshots from example simulations with different mass fraction atmospheres and different impact parameters.  
The trends in atmospheric loss with impact parameters closely follow those identified in previous studies \citep[e.g.,][]{kegerreis_atmospheric_2020,roche_atmospheric_2025} regardless of atmospheric mass fraction, such that loss increases with increasing $v_{\rm c}$ and $\gamma$, but decreases with increasing $b$ (Figure \ref{fig:loss_data}). 
However, there can be significant differences in the absolute efficiency of combined, near, and far-field loss depending on the atmospheric mass fraction. 

\subsection{Dependence of atmospheric loss on initial atmospheric mass}\label{subsec:atmos_mass_dependence}
\begin{figure*}
\begin{center}
    \includegraphics[scale=0.5]{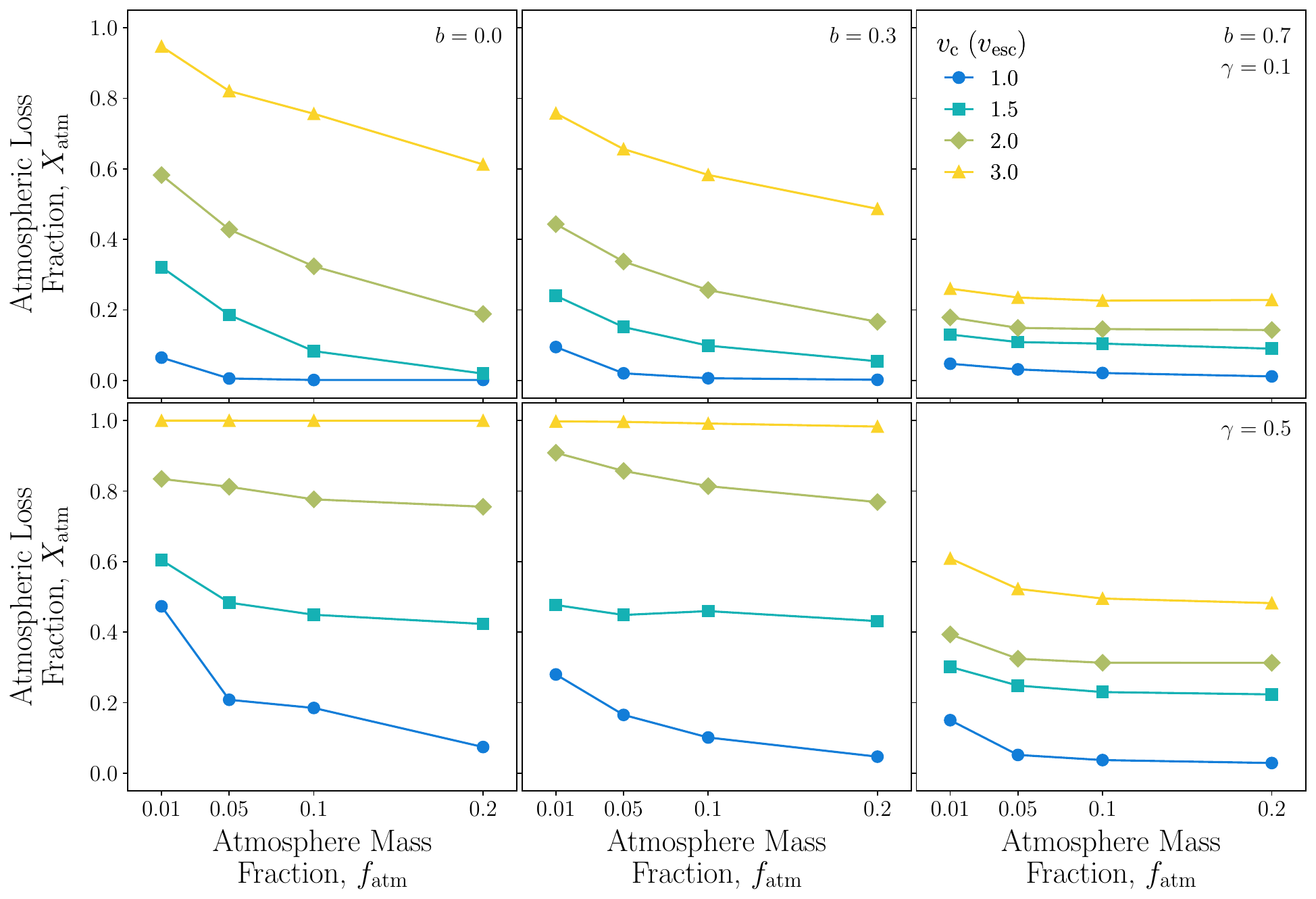}
    \caption{Atmospheric loss increases with decreasing atmosphere mass fraction, but follows previously identified trends with impact angle, velocity, and mass ratio. The data shown are the calculated combined atmospheric loss fractions for impacts with different impact angles/parameters ($b$, columns), impactor-to-total system mass ratios ($\gamma$, rows), and impact velocities ($v_{\rm c}$, colours and symbol types). The near- and far-field components of loss are shown separately in Figures~\ref{fig:X_NF} and \ref{fig:X_FF}. }
    \label{fig:loss_data}
\end{center}
\end{figure*}

We find that the fraction of atmosphere lost is strongly dependent on $f_{\rm atm}$, such that increasingly thin atmospheres are more easily eroded (Figure \ref{fig:loss_data}; see also Table \ref{table:initial_conditions} for a full breakdown).
In some cases, for the same impact parameters (i.e., $v_{\rm c}$--$b$--$\gamma$ combination) there can be an additional 40\% of the original atmosphere lost from a 1\% mass fraction atmosphere compared to a 20\% mass fraction atmosphere.
The trends in loss efficiency with atmospheric mass fraction are dictated by three related effects. 
Firstly, as demonstrated by \citet{lock2024atmospheric}, the efficiency of loss due to the shock-kick mechanism is higher for thinner atmospheres, as the breakout (impedance match) velocity of the impact shock from the mantle into the atmosphere is higher. 
Shock-kick predominantly drives far-field loss \citep{roche_atmospheric_2025} and the increasing breakout velocity is reflected in increasing far-field loss with decreasing $f_{\rm atm}$ (Figure \ref{fig:X_FF}).
Secondly, the fraction of a planet's atmosphere that is in the far field increases with decreasing $f_{\rm atm}$, and loss of a larger fraction of the atmosphere is determined by shock-kick (Figure \ref{fig:X_FF}). 
Finally, in tandem, the amount of atmosphere in the near field decreases with decreasing $f_{\rm atm}$, and the near-field loss saturates for increasingly low energy impacts (Figure \ref{fig:X_NF}). 
As a result, the increasing efficiency of far-field loss becomes the primary driver of the trends in loss with decreasing atmospheric mass. 
The dependence of the combined loss on $f_{\rm atm}$ weakens as impacts become more grazing (increasing $b$, right column of Figure~\ref{fig:loss_data}), since reduced coupling of the impact shock into the target planet suppresses far-field loss, and almost all loss is from the near field which is less strongly affected by the atmospheric mass fraction. 

A final interesting effect emerges for large atmospheric envelopes, where portions of the mid-altitude atmosphere can be shielded from loss. 
This effect can be seen in Figure \ref{fig:impacts} (bottom left), where a lens of atmospheric material initially in the impactor's path is retained (blue particles), whilst surrounding atmosphere is removed (green).
As the impactor enters the atmosphere, the resulting air shock redistributes atmosphere mass away from the impact site, reducing its susceptibility to subsequent erosion. 
This shielding effect lowers the overall loss efficiency, particularly for slower, more head-on impacts.

\subsection{Scaling law for loss} \label{subsec:scaling law}
\begin{figure*}
\begin{center}
    \includegraphics[width=\textwidth]{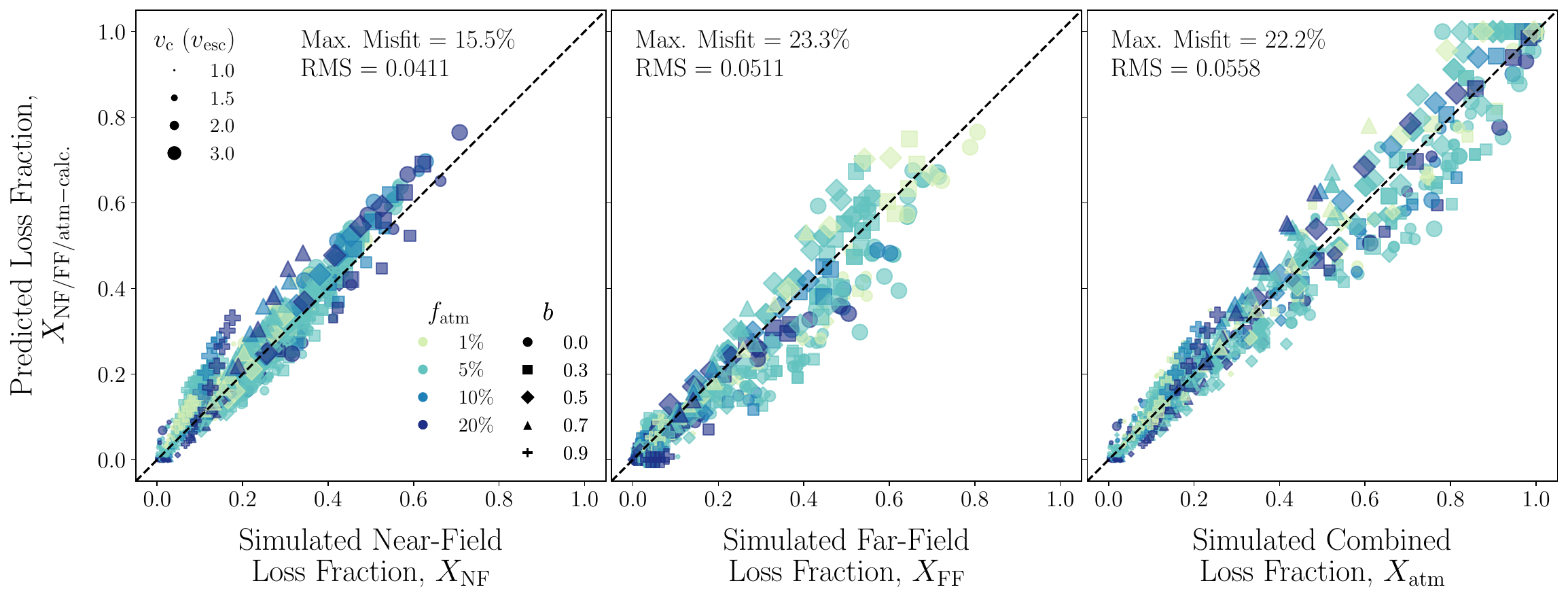}
    \caption{Our new scaling law captures the effects of atmospheric mass on loss and can predict the combined atmospheric loss fraction for the simulations in our fitting dataset with an average misfit in absolute atmospheric loss of $\sim$4.0\%. Simulated (from our SPH simulations) near-field, far-field, and combined loss fraction data (from this study and the data from \citealp{roche_atmospheric_2025}) are plotted against the corresponding values predicted using our scaling law for each value of atmospheric mass fraction (colours), impact parameter (shapes), and impact velocity (marker sizes), with the maximum misfits and rms reported in each panel. Note that we here we do not demarcate different values of $\gamma$ or $M_{\rm t}^{\rm r}$.}
    \label{fig:scaling_law}
\end{center}
\end{figure*}

Here, we expand the scaling law for atmospheric loss of \citet{roche_atmospheric_2025} to account for the effect of atmospheric mass. 
The combined atmospheric loss fraction, $X_{\rm atm}$ is calculated as a sum of its near- and far-field components, $X_{\rm NF}$ and $X_{\rm FF}$, respectively:

\begin{equation}\label{eq:X_atm}
    X_{\rm atm} = X_{\rm NF} + X_{\rm FF} \; .
\end{equation}

\noindent We use the same functional forms for the near- and far-field loss as in \citet{roche_atmospheric_2025}, but now include additional terms to capture the effects of changing atmosphere mass and a pre-factor to explicitly account for the fraction of the atmosphere in the near/far field, which varies significantly with atmospheric mass fraction and impact parameters. 
The fraction of atmosphere in the near field,  $f_{\rm NF}$, was calculated using the tangent surface method \citep{roche_atmospheric_2025}.
However, calculations using this method are too computationally expensive for the purposes of a scaling law. 
We therefore fit a functional form to a suite of initialised SPH planets with impactor-to-target radius ratios $R^{\rm r}_{\rm i}/R^{\rm r}_{\rm t}=$ 0.1--1.0 (sampled at intervals of 0.1, with an additional point at 0.001 to constrain the fit for smaller bodies), $M_{\rm t}^{\rm r} \in$ \{0.01, 0.05, 0.1, 0.35, 1.0, 2.0, 5.0\} M$_{\oplus}$, and $f_{\rm atm} \in$ \{0.01, 0.05, 0.1, 0.2\}.
The impact parameter was sampled using a Chebyshev distribution,
\begin{equation}
    b = \frac{1}{2}\left[1 + \cos\!\left(\frac{\pi(2j - 1)}{50}\right)\right], \quad j = 1, 2, \dots, 25 \;,
\end{equation}

\noindent ensure stability close to $b = 0$ and $b = 1$.
We find that $f_{\rm NF}$ is well-captured by a generalised logistics function:

\begin{equation}\label{eq:f_NF}
    f_{\rm NF} = \frac{\zeta_4}{1 + \zeta_6\exp\left(\zeta_3\left(\frac{R_{\rm i}^{\rm r}}{R_{\rm t}^{\rm r}}\right) - \zeta_2 \right)^{\zeta_1}} + \zeta_5 \;,
\end{equation}

\noindent where $\zeta_i$ are functions of $b$, $f_{\rm atm}$, and $M_{\rm t}^{\rm r}$:

\begin{equation}\label{eq:zeta_i}
\begin{split}
    \zeta_{i\in\{1-4\}} &= q_{i1} + q_{i2}b + q_{i3}b^{q_{i4}} + q_{i5}f_{\rm atm} \\ & + q_{i6}f_{\rm atm}^2 + q_{i7}\left(\frac{M_{\rm t}^{\rm r}}{M_{\oplus}} \right)^{q_{i8}} \;,
\end{split}
\end{equation}

\noindent and $q_{ij}$ are fitted parameters where $i$ refers to the relevant $\zeta_i$ and $j$ to the $j^{\rm th}$ fitting parameter; $\zeta_5$ and $\zeta_6$ are single values (Table \ref{table:f_NF_params}).
Our function accurately fits the synthetic data with mean and maximum misfits of $\sim$3.1\% and $\sim$9.4\%, respectively.

Using this expression for $f_{\rm NF}$, the near-field loss is given by

\begin{equation}\label{eq:X_NF}
    X_{\rm NF} = f_{\rm NF} \cdot \left(\xi_1 - \xi_2\left(b + \xi_3\right)^2 \right) \; ,
\end{equation}

\noindent where $\xi_i$ are functions of $\gamma$, $v_{\rm c}$, $M_{\rm t}^{\rm r}$, and $f_{\rm atm}$:

\begin{equation}\label{eq:xi_i}
\begin{split}
    \xi_{i\in\{1-3\}} &= k_{i1} + k_{i2}\gamma + k_{i3}\left(\gamma+0.05\right)^2 + k_{i4}\left(\frac{v_{\rm c}}{v_{\rm esc}}\right)^{k_{i5}} \\ & 
    + k_{i6}\left(\frac{M_{\rm t}^{\rm r}+1.0}{M_{\oplus}}\right) + k_{i7}\log_{10}\left(f_{\rm atm}\right) \; .
\end{split}
\end{equation}

\noindent $k_{ij}$ are fitted parameters where $i$ refers to the relevant $\xi_i$ and $j$ to the $j^{\rm th}$ fitting parameter (Table \ref{table:NF_params}). This functional form was chosen empirically as giving a good fit to the loss results \citep{roche_atmospheric_2025}.
The constants 0.05 and 1.0 in the third and fifth terms, respectively, have been added to ensure each term tends towards a finite value for small $\gamma$ and $M_{\rm t}^{\rm r}$, allowing for stable extrapolation.

The far-field loss is given by

\begin{equation}\label{eq:X_FF}
    X_{\rm FF} = f_{\rm FF} \cdot \left(\psi_1 \exp\left[-\psi_2\,Q_{\rm R}^{\prime}\left(1 + \frac{M_{\rm i}^{\rm r}}{M_{\rm t}^{\rm tot}}\right)\left(1 - b\right)^{\psi_4}\right] + \psi_3\right) \;,
\end{equation} 

\noindent where $\psi_i$ are functions of $\gamma$, $M_{\rm t}^{\rm r}$, and $f_{\rm atm}$:

\begin{equation}\label{eq:psi_i}
\begin{split}
    \psi_{i \in \{1-4\}} &= s_{i1} + s_{i2}\,\left(\gamma+0.05\right)^{s_{i3}} + s_{i4}\left(\frac{M_{\rm t}^{\rm r}+0.05}{M_{\oplus}}\right)^{s_{i5}} \\ &+ s_{i6}\log_{10}\left(f_{\rm atm}\right) \;.
\end{split}
\end{equation}

\noindent $s_{ij}$ are fitted parameters where $i$ refers to the relevant $\psi_i$ and $j$ to the $j^{\rm th}$ fitting parameter (Table \ref{table:FF_params}).
The functional form for $X_{\rm FF}$ is physically motivated to capture the dependence of the strength of the impact shock on impact parameters \citep{roche_atmospheric_2025}.
The 0.05 in the third and fourth terms ensure each tends towards a finite value for small $\gamma$ and $M_{\rm t}^{\rm r}$, allowing stable extrapolation.
$f_{\rm FF}=1 - f_{\rm NF}$ is the fraction of atmosphere mass in the far field.

To constrain the parameters for the scaling law, we used a linear least squares fit (implemented using the \texttt{leastsq} function from the \texttt{scipy.optimize} package, version 1.9.1), minimising the cost function 

\begin{equation}
    J=\sum_{i}{\left( \frac{y_{\rm model}^i - y_{\rm data}^i}{|y_{\rm data}^i| + 0.1}\right)^2} \; ,
\end{equation} 

\noindent between the model and the results from this study and \citet{roche_atmospheric_2025}. 
Inclusion of the latter, allowed us to constrain the dependence on target planet mass since the simulations run for this study all had a single refractory target mass.
Using a modified cost function, rather than a classic linear-least squares fit, allowed us to better balance the fit between the larger absolute differences for high loss values and larger fractional differences for very low loss values.

Our new scaling law accurately predicts atmospheric loss over the wide range of impacts and atmosphere masses in our simulation set, with mean and maximum misfits of $\sim$4.0\% and $\sim$22.2\%, respectively (Figure \ref{fig:scaling_law}).
A new \texttt{python} function for calculating the atmospheric loss for any given impact scenario is available on GitHub (\url{https://github.com/mjr230600/Roche_et_al_2026_ApJL_scaling-law_plotting-scripts.git}).

\section{Discussion} \label{sec:discussion}

\subsection{Caveats for use of our scaling law}
\label{subsec:caveats}

Our scaling law should be used with caution outside of the range constrained by our simulation suite and that of \citet{roche_atmospheric_2025} i.e., $M_{\rm t}^{\rm r}$ = 0.35--5.0 M$_{\oplus}$, $f_{\rm atm}$ = 0.01--0.2, $\gamma$ = 0.1--0.5, $b$ = 0.0--0.9, and $v_{\rm c}=$ 1.0--3.0\,$v_{\rm esc}$. 
However, to support its use in coupled accretion models, we verify that our scaling law remains numerically stable to extrapolation to $M_{\rm t}^{\rm r} = 10^{-3}$--10 M$_{\oplus}$, $\gamma=10^{-3}$--0.5, and $f_{\rm atm}=10^{-6}$--$0.4$, although accuracy is not guaranteed beyond the fitted range.

The scaling law also does not yet include terms accounting for planetary rotation or other factors known to affect atmospheric loss, such as atmospheric composition, temperature, or presence of surface oceans \citep[][M. J. Roche et al., in review]{genda2005enhanced,lock2024atmospheric}.
Additionally, the thick volatile envelopes of sub-Neptune planets may be miscible with their silicate interiors \citep[e.g.,][]{young_differentiation_2025,rogers_redefining_2025,gilmore_coreenvelope_2026}, potentially shielding volatiles from impact-driven loss. 
Since our simulations assumed a distinct atmosphere--mantle boundary, atmospheric loss may be overestimated for larger, particularly young, planets with massive atmospheres.

Further, our scaling law does not account for the effects of thermally-driven atmospheric loss following a giant impact, which, given the extreme heating of underlying core and mantle material during a giant impact \citep{postema_thermal_2026}, has been proposed to be an important secondary loss mechanism for primordial envelopes \citep{biersteker2021losing}.
Post-impact structures are largely optically thick mixtures of vapour, liquid, and supercritical fluid \citep{lock2018origin} with mean-free-paths much less than a meter, except for the outermost low-density regions.
Energy transfer is therefore dominantly governed by advective flows rather than radiation, and we therefore do not expect radiative processes to substantially affect the extent of loss, certainly within the first 10's of hours after impact as considered here where mechanical loss mechanisms dominate.
\citet{biersteker2021losing} proposed that heating from the silicate core of a post-impact body could drive a hydrodynamic outflow and significant additional atmospheric loss for primordial H$_2$--He envelopes, an effect they show becomes negligible for higher mean molecular weight envelopes.
This secondary mechanism for atmosphere removal would follow potentially some many hundreds of years \citep[][]{lock_energy_2020} after the immediate shock-driven loss of the atmosphere calculated here and in \citet{roche_atmospheric_2025}. 
However, we stress that the connection between the post-giant-impact state and the initial conditions assumed in post-impact-driven loss models is yet to be explored, and there could be a significant delay between the mechanical ejection of atmosphere considered here and such secondary loss mechanisms.
We can, however, comment that since our scaling laws only consider the immediate effects of shock- and vapour-plume driven atmospheric loss, and neglect later thermally-driven processes relevant for primordial envelopes, our scaling law predictions could be lower limit estimates for the final extent of impact-driven atmospheric loss.
Therefore, despite still being considerably more generalised than previous prescriptions \citep[e.g.,][]{kegerreis_atmospheric_2020,denman2022atmosphere,roche_atmospheric_2025}, our new scaling law must be used with these caveats in mind, and could underestimate the final extent of impact-driven loss when considering secondary loss mechanisms \citep[e.g.,][]{biersteker2021losing}.

\subsection{Impact-driven atmospheric loss during planet formation}\label{subsec:N_body}

\begin{figure*}
\begin{center}
    \includegraphics[scale=0.6]{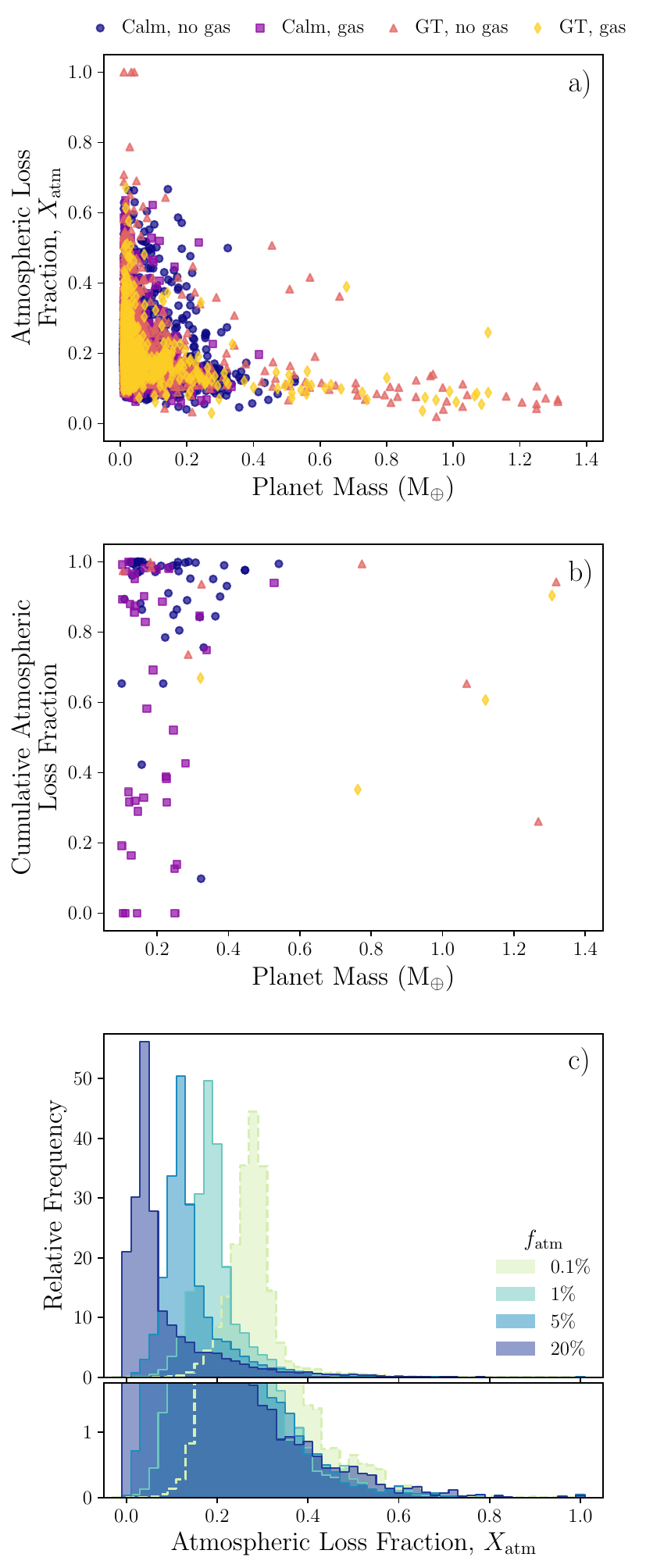}
    \caption{The cumulative effect of many giant impacts during planet formation can lead to a large variations in the volatile budgets of planetary bodies, including total loss of atmosphere. Shown are: a) individual loss events as a function of planet mass from planets with $f_{\rm atm}$ = 1\% only; b) the cumulative losses resulting from multiple impacts throughout accretion for different formation scenarios (symbols and colours); and c) a histogram of loss efficiency from all individual impacts in the $N$-body accretion simulations calculated using our scaling law, assuming different pre-impact atmosphere mass fractions, $f_{\rm atm}$ (colours). The relative frequency of impacts has been scaled by the number of impacts in each scenario so as not to weight any one scenario preferentially, and the total area under each histogram equals four, the number of scenarios considered. The lower panel is zoomed in on the lower frequency events. The loss for $f_{\rm atm}$ = 0.1\% (dashed outline) is an extrapolation of our scaling law beyond the simulation data used to fit it.
    }
    \label{fig:N_body}
\end{center}
\end{figure*}

To assess the cumulative effect of atmospheric loss throughout planet formation, and the likelihood of significant loss from individual events, we applied our new scaling law to the giant impacts from 14 existing $N$-body simulations of four separate solar system formation scenarios from \citet{carter2015compositional} and \citet{carter2022did}, similarly to \citet{roche_atmospheric_2025}.
These comprise two end-member dynamical scenarios for the inner solar system, both simulated with and without nebular gas: calm disk, which neglects giant planet perturbations; and Grand Tack, in which Jupiter migrates inward and then outward, truncating and exciting the inner disk \citep{walsh2011low}.
We applied our atmospheric erosion scaling law to all collisions with refractory target masses $\geq$0.01\,M$_\oplus$ and impactor-to-target mass ratios $\geq$0.01, following the definition of `giant impacts' as events involving significant changes in potential energy \citep{carter2020energy}. 
We quantify the cumulative effects of impact-driven atmospheric loss using the `relative erosion' metric ranging from 0 (no loss) to 1 (complete loss), calculating the ratio of the results between cases where the scaling law is applied to all impacts and equivalent cases with no erosion.
This isolates the effects of impact-driven atmospheric removal from changes due to refractory accretion and volatile delivery.
Bodies that reach a minimum mass of 0.01\,M$_\oplus$ within the lifetime of the nebular gas (set at 5\,Myr) acquire an atmosphere according to Equation 13 from \citet{mordasini_planetary_2020}, which is derived from the analytical model of \citet{lee_cool_2015}, assuming a semi-major axis of 1\,au. 
We calculate the effect of each collision on the composition of the remnants, removing atmosphere according to our scaling law. 
Second largest remnants, if produced from a collision, are assigned atmosphere from the projectile (partial accretion and hit-and-run collisions) or the remaining portion of the target (erosive and super catastrophic collisions) proportionally to its mass. 
The atmosphere masses of target and projectile in each subsequent collision are taken from the remnants of the previous collision each body was involved in assuming all volatiles are within the atmosphere pre-impact. 
The results of this analysis are presented in Figure \ref{fig:N_body}.

We find that the cumulative effect of many giant impacts during planet formation leads to a wide range in atmospheric loss across all planet masses due to the stochastic nature of accretion (Figure \ref{fig:N_body}a,b). 
Planets with final masses $\lesssim$0.6 M$_{\oplus}$ typically lose $>$50\% of their atmospheres, whilst roughly Earth-mass planets ($\gtrsim$0.6 M$_{\oplus}$) display highly variable losses of $\sim$30--100\%. 
As expected, the median loss per impact increases as atmospheric mass fraction decreases (Figure \ref{fig:N_body}c): 20\% mass fraction atmospheres typically lose 0--5\%, 1\% atmospheres $\sim$20\%, and 0.1\% atmospheres (extrapolated using our scaling law) lose $\sim$30\%.
Individual high-loss events ($>$50\%) are rare, particularly for impacts onto Earth-mass planets (Figure \ref{fig:N_body}a,b) or those with $>$0.1\% mass fraction atmospheres (Figure \ref{fig:N_body}c).
Larger planets that experienced significant cumulative atmospheric loss did so by accreting embryos that experienced large degrees of loss and/or by undergoing multiple atmospheric-stripping impacts.

Our results suggest that impact-driven atmospheric loss during planet formation is more efficient than previously thought. 
The average cumulative loss from our calculations is greater than that calculated by \citet{roche_atmospheric_2025}, since they did not consider the effect of varying atmospheric mass fraction and applied a scaling law for fixed 5\% mass fraction atmospheres to all impacts. 
For small bodies that initially accrete lower mass envelopes, loss in every impact is more efficient, leading to a much greater cumulative loss.
Consequently, we find that many smaller bodies experience total atmospheric loss when there were no such cases in \citet{roche_atmospheric_2025}. 
There is also a feedback between atmospheric loss in one event decreasing the atmospheric mass fraction and so the susceptibility to loss in subsequent events. 
The impact of this feedback is seen in some larger bodies where the maximum loss jumps from $\sim$80\% in the calculations of \citet{roche_atmospheric_2025} to total loss here, as some later impacts onto bodies that had already experienced significant loss were much more efficient at removing atmosphere.
In reality, this feedback would not be as straightforward as in our calculations as outgassing and/or volatile delivery, not considered here, could also alter the mass of the atmosphere between impacts.

An increased prevalence of atmospheric loss during accretion could help to explain the range of super-solar $^3$He/$^{22}$Ne in Earth's mantle. 
\citet{tucker2014evidence} argue that this signature requires two or more highly-efficient atmosphere stripping events capable of removing most of Earth's atmosphere whilst simultaneously causing substantial mantle melting. 
Our results indicate that substantial atmospheric loss from planetary embryos is common.
Moreover, late-stage impacts onto the proto-Earth could have removed large fractions of its atmosphere if they were sufficiently energetic \citep[as suggested by recent Moon-formation models;][]{lock2018origin,cuk_making_2012,canup_forming_2012} or if the pre-impact atmosphere was relatively thin. 
Our study therefore gives mechanistic feasibility to the proposal by \citet{tucker2014evidence}. 

More broadly, our results indicate that in systems that undergo a phase of giant impacts, retention of accreted nebular atmospheres is a matter of chance. 
A substantial fraction of atmospheres observed around rocky exoplanets are most likely to be secondary in nature and derived from rainout of post-impact silicate vapour atmospheres \citep[e.g.,][]{caracas2023no}, outgassing during mantle crystallisation \citep[e.g.,][]{nicholls_magma_2024}, or long-term volcanic degassing \citep[e.g.,][]{ortenzi_mantle_2020}, even if they had a nebular atmosphere at the point of nebula dispersal. 
It is also very likely that planets in the same system will have atmospheres with different origins due to stochastic differences in their accretion \citep[e.g.,][]{kite_exoplanet_2020,krissansen-totton_erosion_2024}.
Although our calculations neglect several potentially important processes (i.e., volatile delivery by the impactor, atmosphere--interior re-equilibration between impacts, evolving atmospheric compositions or the presence of surface oceans, and cumulative erosion by smaller impactors, \citealp[e.g.,][]{schlichting_atmospheric_2015}) that require more generalised scaling laws and development of coupled geochemical/dynamical models, our results nevertheless demonstrate that giant impacts are a critical and inherently stochastic control on planetary volatile budgets.
Incorporating our new scaling law into larger-scale frameworks of whole planet and system-wide evolution \citep[e.g.,][]{gu2024composition,nicholls_magma_2024} will likely provide further insight into the atmospheric diversity of terrestrial planets.

\subsection{Atmospheric loss during the Moon-forming impact}\label{subsec:MFI}
\begin{figure*}
\begin{center}
    \includegraphics[scale=0.6]{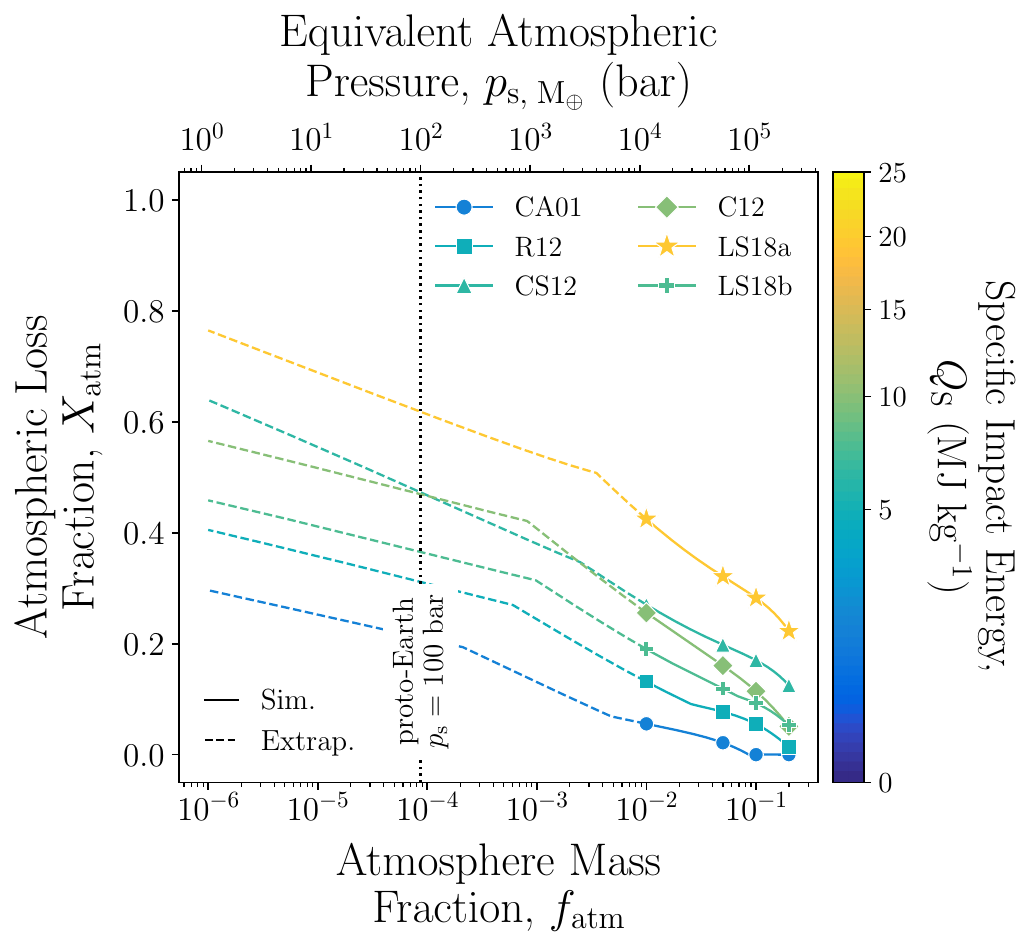}
    \caption{Earth could have lost significant portions of its atmosphere during the Moon-forming giant impact with significant variation depending on the impact scenario and initial atmospheric mass. Atmospheric loss fraction as a function of atmosphere mass fraction predicted by our scaling law for six Moon-formation scenarios (Table \ref{tab:MFI_scenarios}). Values of $f_{\rm atm}$ within the range simulated in this study are shown as solid lines (Sim.) and lower value extrapolations as dashed lines (Extrap.), with line colours indicating the specific impact energy. Note, our calculation does not include rotation which could significantly alter the loss efficiency in the \citet{cuk_making_2012} case. Equivalent atmospheric pressures shown on the upper $x$-axis were calculated for a 1.0 M$_{\oplus}$ planet and thus serve only as a reference for the data points shown here. The illustrative atmospheric mass fraction for the proto-Earth, corresponding to an atmospheric pressure on an Earth-mass body of 100~bar \citep{zahnle_emergence_2007,sossi_redox_2020}, discussed in the text is shown as a black dotted vertical line. The inflection points seen in many scenarios marks the point of near-field loss saturation, with additional loss coming only from the far field.
    }
    \label{fig:MFI}
\end{center}
\end{figure*}

The Moon-forming giant impact (MFI) is thought to be Earth's last major accretionary event \citep{hartmann_satellite-sized_1975,cameron_and_ward_origin_1976} and set the stage for Earth's subsequent evolution \citep[e.g.,][]{lock_geochemical_2020}.
Atmospheric loss during Moon formation therefore, in part, dictates what fraction of Earth's volatiles were inherited from the main stage of accretion vs. delivered afterwards during late accretion.
This has significant implications for the state of the early atmosphere and our interpretation of geochemical tracers of accretion.

Prior works have attempted to predict the extent of atmospheric loss experienced during the MFI, either using shock-driven ground motion calculations \citep[e.g.,][]{genda2003survival,genda2005enhanced,schlichting_atmospheric_2015,yalinewich2019atmospheric} or SPH-derived scaling laws \citep{kegerreis_atmospheric_2020}, and arrived at values ranging from $\sim$5--60\%.
However, these previous attempts either did not include all impact-driven mechanisms of atmospheric loss or did not consider the effect of the pre-impact atmospheric mass on the degree of loss.

Using our new scaling law, we estimate how much of the proto-Earth's atmosphere could have been lost during the MFI as a function of initial atmospheric mass in six different Moon formation scenarios (Table \ref{tab:MFI_scenarios}): `canonical' \citep[][CA01]{canup_origin_2001}, `hit-and-run' \citep[][R12]{reufer_hit-and-run_2012}, and four higher-energy synestia-forming impacts including the originally proposed scenarios of a rapidly-spinning proto-Earth \citep[][CS12]{cuk_making_2012}, the near-symmetric impact of two half-Earths \citep[][C12]{canup_forming_2012}, and two additional impacts scenarios from \citet[][]{lock2018origin} (LS18a,b). 
The results of this analysis are shown in Figure~\ref{fig:MFI}. 
Note that our calculation does not include the effects of rotation which could significantly alter the loss efficiency in the \citet{cuk_making_2012} case. 

We find that the MFI could have removed a wide range of the proto-Earth's atmosphere, depending both on the style of impact and on the mass of the pre-impact atmosphere, with the two effects having comparable magnitudes.
Taking the proto-Earth's pre-impact atmosphere as $\leq$100~bar \citep[e.g.,][]{zahnle_earths_2010,sossi_redox_2020}, i.e., $f_{\rm atm}\approx10^{-4}$ for a 1.0 M$_{\oplus}$ planet, we find that the least energetic scenario, the `canonical MFI' \citep{canup_origin_2001}, could have removed between $\sim$20--30\% of the proto-Earth's atmosphere.
In contrast, the most energetic scenario, a synestia-forming scenario from \citet{lock2018origin} similar to the two half-Earths model from \citet{canup_forming_2012}, could have removed most of the proto-Earth's atmosphere, between $\sim$70--80\% -- 10's of percent higher than the previous upper limit estimates of $<$60\% proposed by \citet{kegerreis_atmospheric_2020}. 
Therefore, it is possible to impose a combined constraint on Earth's volatile budget at the time of Moon formation, how that inventory was partitioned between Earth's interior and atmosphere, and the scenario of Moon formation.
Future models that trace volatile evolution through accretion (akin to \citealp{gu2024composition}) may therefore be able to apply this constraint on, or at least describe tradeoffs with, the style of the MFI.

To caveat these estimates, our scaling law was derived using H$_2$--He atmospheres, whereas the heavier CO- or CO$_2$-rich atmosphere likely present on the proto-Earth \citep{zahnle_earths_2010,sossi_redox_2020} would be less susceptible to shock-driven loss \citep{lock2024atmospheric}, making our estimates upper limits for a given atmosphere mass. 
We also neglect the effects of surface oceans, which can either enhance or suppress atmospheric loss depending on the atmosphere-to-ocean mass ratio \citep[][M. J. Roche et al., in review]{genda2005enhanced,lock2024atmospheric}, as well as planetary rotation, which may increase loss efficiency for rapidly rotating pre-impact bodies. 
These effects will be explored in future work.

\section{Conclusions} \label{sec:conclusions}

We used a large suite of new and existing SPH giant impact simulations with atmosphere mass fractions from 20\% to 1\% to show that increasingly thin atmospheres are more easily removed during impacts.
We developed a new scaling law to account for the effects of changing atmosphere mass that is applicable across a wide range of planet and atmosphere masses, impact angles, velocities, and impactor-to-total system mass ratios, although caution is advised when extrapolating beyond the fitted regime.

Based on our results, we show that impact-driven atmospheric loss is likely more efficient during accretion than previously thought, due to the dependence of atmospheric loss on atmospheric mass and the resulting feedback between previous loss and enhancement of loss in subsequent impacts. 
Applying our scaling law to $N$-body simulations for solar system formation revealed that planets $\lesssim$0.6 M$_{\oplus}$ often lose their entire atmospheres during accretion, whilst larger bodies can experience between 30\% and total loss. 
The prevalence of impact-driven atmospheric loss during the accretion of Earth-mass planets would have had a significant effect on the atmospheres of the terrestrial planets, and could explain the variation in $^3$He/$^{22}$Ne found in Earth's mantle \citep{tucker2014evidence}.
Further, we showed that the amount of the proto-Earth's atmosphere that survived Moon formation is strongly dependent on both the pre-impact atmosphere mass fraction and the impact scenario itself. 
Earth would have lost between 20--80\% of its atmosphere if it had a $\sim$100 bar atmosphere at the time of impact.

Atmospheric loss due to giant impacts likely plays an important role in shaping the atmospheres of planets in any systems that undergo a phase of giant impacts. 
Incorporating our scaling laws for impact-driven atmospheric loss into larger-scale frameworks of planet formation and evolution \citep[e.g.,][]{gu2024composition,nicholls_magma_2024} will elucidate the links between accretion dynamics and the diversity of planetary volatile budgets. 

\begin{acknowledgments}
\noindent MJR acknowledges financial support from the STFC, UK (grant number: ST/X508251/1). SJL acknowledges financial support from NERC, UK (grant number: NE/V014129/1). ZML and PJC acknowledge financial support from the STFC, UK (grant numbers: ST/V000454/1 and ST/Y002024/1). The authors declare that they have no competing interests. We thank the anonymous reviewer for their feedback in improving our manuscript. 
\end{acknowledgments}

\begin{contribution}
Both MJR and SJL devised the project. MJR ran and analysed 3D SPH simulations, developed the scaling law, performed analysis of atmospheric loss during the Moon-forming impact, and wrote the text. SJL discussed results, and edited the text. PJC conducted analysis of $N$-body simulations, discussed results, and edited the text. ZML discussed results and edited the text. 

\end{contribution}

%
\facilities{All simulations were carried out using the computational facilities of the Advanced Computing Research Centre, University of Bristol – \url{http://www.bristol.ac.uk/acrc/}.}

\software{This work made use of \texttt{SWIFT} \citep{schaller_swift_2024, kegerreis2019planetary}, \texttt{WoMa} \citep{ruiz2021effect}, \texttt{SEAGen} \citep{kegerreis2019planetary}, \texttt{numpy} \citep{van2011numpy}, \texttt{matplotlib} \citep{hunter2007matplotlib}, \texttt{scipy} \citep{virtanen2020scipy}, \texttt{pandas} \citep{mckinney2011pandas}, and \texttt{jupyter} \citep{kluyver2016jupyter}}.


\appendix
\renewcommand{\thefigure}{B\arabic{figure}}   
A \texttt{python} function of our atmospheric loss scaling law, and code required to reproduce the figures and results from this paper, can be found on GitHub (\url{https://github.com/mjr230600/Roche_et_al_2026_ApJL_scaling-law_plotting-scripts.git}) and Zenodo, which will be made available upon acceptance. 
Initial conditions files for our simulation suite are available on request.

\section{Impact scenarios}\label{sec:impact_scenarios}
The initial conditions for each simulation along with the resultant near-field, far-field, and combined atmospheric losses are listed below in Table \ref{table:initial_conditions}.

\renewcommand{\thetable}{\thesection\arabic{table}}
\begin{longtable*}{ccccccccccccc}
    \caption{Planet masses ($M_{\rm t}^{\rm tot}$, $M_{\rm i}^{\rm r}$) and radii (including atmosphere, for target and impactor; $R_{\rm t}^{\rm tot}$, $R_{\rm i}^{\rm r}$), atmosphere mass fractions ($f_{\rm atm}$), mass ratios ($\gamma$), impact velocities ($v_{\rm c}$), impact parameters ($b$), and resultant near-field ($X_{\rm NF}$), far-field ($X_{\rm FF}$), and combined ($X_{\rm atm}$) atmospheric losses for each impact scenario performed in this study. A machine readable form of this table is available in the Supplement.} \\
    \toprule
    \toprule
    $M_{\rm t}^{\rm tot}$ (M$_{\oplus}$) & $R_{\rm t}^{\rm tot}$ (R$_{\oplus}$) & $M_{\rm i}^{\rm r}$ (M$_{\oplus}$) & $R_{\rm i}^{\rm r}$ (R$_{\oplus}$) & $f_{\rm atm}$ & $\gamma$ $\left(M_{\rm i}^{\rm r}/M_{\rm tot}^{\rm r}\right)$ & $v_{\rm c}$ $\left(\text{km s}^{-1}\right)$ & $v_{\rm c}$ ($v_{\rm esc}$) & $b$ & $\theta$ $\left(^\circ\right)$ & $X_{\rm NF}$ & $X_{\rm FF}$ & $X_{\rm atm}$ \\
    \midrule
    \endfirsthead

    \toprule
    \toprule
    $M_{\rm t}^{\rm tot}$ (M$_{\oplus}$) & $R_{\rm t}^{\rm tot}$ (R$_{\oplus}$) & $M_{\rm i}^{\rm r}$ (M$_{\oplus}$) & $R_{\rm i}^{\rm r}$ (R$_{\oplus}$) & $f_{\rm atm}$  & $\gamma$ $\left(M_{\rm i}^{\rm r}/M_{\rm tot}^{\rm r}\right)$ & $v_{\rm c}$ $\left(\text{km s}^{-1}\right)$ & $v_{\rm c}$ ($v_{\rm esc}$) & $b$ & $\theta$ $\left(^\circ\right)$ & $X_{\rm NF}$ & $X_{\rm FF}$ & $X_{\rm atm}$ \\
    \midrule
    \endhead

    \bottomrule
    \endfoot
    
    \begin{filecontents*}{impact_scenarios.csv}
M_t_tot,R_t_tot,M_i,R_i,f_atm,gamma,v_c,v_c_v_esc,b,theta,X_NF,X_FF,X_atm
1.007,1.181,0.111,0.524,0.01,0.1,9.53,1.0,0.0,0.0,0.037,0.028,0.064
1.007,1.181,0.111,0.524,0.01,0.1,9.53,1.0,0.3,17.5,0.058,0.036,0.094
1.007,1.181,0.111,0.524,0.01,0.1,9.53,1.0,0.5,30.0,0.058,0.012,0.070
1.007,1.181,0.111,0.524,0.01,0.1,9.53,1.0,0.7,44.4,0.041,0.006,0.047
1.007,1.181,0.111,0.524,0.01,0.1,9.53,1.0,0.9,64.2,0.019,0.014,0.033
1.007,1.181,0.111,0.524,0.01,0.1,14.29,1.5,0.0,0.0,0.121,0.200,0.321
1.007,1.181,0.111,0.524,0.01,0.1,14.29,1.5,0.3,17.5,0.133,0.107,0.240
1.007,1.181,0.111,0.524,0.01,0.1,14.29,1.5,0.5,30.0,0.136,0.058,0.194
1.007,1.181,0.111,0.524,0.01,0.1,14.29,1.5,0.7,44.4,0.090,0.040,0.130
1.007,1.181,0.111,0.524,0.01,0.1,14.29,1.5,0.9,64.2,0.039,0.018,0.057
1.007,1.181,0.111,0.524,0.01,0.1,19.05,2.0,0.0,0.0,0.143,0.440,0.582
1.007,1.181,0.111,0.524,0.01,0.1,19.05,2.0,0.3,17.5,0.152,0.291,0.443
1.007,1.181,0.111,0.524,0.01,0.1,19.05,2.0,0.5,30.0,0.166,0.165,0.331
1.007,1.181,0.111,0.524,0.01,0.1,19.05,2.0,0.7,44.4,0.113,0.066,0.178
1.007,1.181,0.111,0.524,0.01,0.1,19.05,2.0,0.9,64.2,0.047,0.023,0.069
1.007,1.181,0.111,0.524,0.01,0.1,28.58,3.0,0.0,0.0,0.141,0.806,0.947
1.007,1.181,0.111,0.524,0.01,0.1,28.58,3.0,0.3,17.5,0.161,0.596,0.757
1.007,1.181,0.111,0.524,0.01,0.1,28.58,3.0,0.5,30.0,0.180,0.364,0.544
1.007,1.181,0.111,0.524,0.01,0.1,28.58,3.0,0.7,44.4,0.132,0.128,0.260
1.007,1.181,0.111,0.524,0.01,0.1,28.58,3.0,0.9,64.2,0.058,0.039,0.097
1.007,1.181,0.249,0.675,0.01,0.2,9.64,1.0,0.0,0.0,0.097,0.015,0.112
1.007,1.181,0.249,0.675,0.01,0.2,9.64,1.0,0.3,17.5,0.106,0.071,0.177
1.007,1.181,0.249,0.675,0.01,0.2,9.64,1.0,0.5,30.0,0.075,0.022,0.097
1.007,1.181,0.249,0.675,0.01,0.2,9.64,1.0,0.7,44.4,0.060,0.030,0.090
1.007,1.181,0.249,0.675,0.01,0.2,9.64,1.0,0.9,64.2,0.022,0.014,0.036
1.007,1.181,0.249,0.675,0.01,0.2,14.46,1.5,0.0,0.0,0.209,0.249,0.458
1.007,1.181,0.249,0.675,0.01,0.2,14.46,1.5,0.3,17.5,0.232,0.156,0.388
1.007,1.181,0.249,0.675,0.01,0.2,14.46,1.5,0.5,30.0,0.176,0.072,0.248
1.007,1.181,0.249,0.675,0.01,0.2,14.46,1.5,0.7,44.4,0.113,0.057,0.170
1.007,1.181,0.249,0.675,0.01,0.2,14.46,1.5,0.9,64.2,0.046,0.023,0.070
1.007,1.181,0.249,0.675,0.01,0.2,19.27,2.0,0.0,0.0,0.209,0.547,0.756
1.007,1.181,0.249,0.675,0.01,0.2,19.27,2.0,0.3,17.5,0.254,0.366,0.619
1.007,1.181,0.249,0.675,0.01,0.2,19.27,2.0,0.5,30.0,0.208,0.234,0.442
1.007,1.181,0.249,0.675,0.01,0.2,19.27,2.0,0.7,44.4,0.137,0.085,0.222
1.007,1.181,0.249,0.675,0.01,0.2,19.27,2.0,0.9,64.2,0.056,0.028,0.084
1.007,1.181,0.249,0.675,0.01,0.2,28.91,3.0,0.0,0.0,0.207,0.789,0.997
1.007,1.181,0.249,0.675,0.01,0.2,28.91,3.0,0.3,17.5,0.258,0.647,0.905
1.007,1.181,0.249,0.675,0.01,0.2,28.91,3.0,0.5,30.0,0.221,0.461,0.681
1.007,1.181,0.249,0.675,0.01,0.2,28.91,3.0,0.7,44.4,0.155,0.191,0.346
1.007,1.181,0.249,0.675,0.01,0.2,28.91,3.0,0.9,64.2,0.067,0.044,0.110
1.007,1.181,0.428,0.796,0.01,0.3,9.95,1.0,0.0,0.0,0.152,0.023,0.174
1.007,1.181,0.428,0.796,0.01,0.3,9.95,1.0,0.3,17.5,0.122,0.057,0.179
1.007,1.181,0.428,0.796,0.01,0.3,9.95,1.0,0.5,30.0,0.098,0.031,0.129
1.007,1.181,0.428,0.796,0.01,0.3,9.95,1.0,0.7,44.4,0.064,0.023,0.087
1.007,1.181,0.428,0.796,0.01,0.3,9.95,1.0,0.9,64.2,0.025,0.012,0.036
1.007,1.181,0.428,0.796,0.01,0.3,14.93,1.5,0.0,0.0,0.275,0.389,0.663
1.007,1.181,0.428,0.796,0.01,0.3,14.93,1.5,0.3,17.5,0.279,0.164,0.443
1.007,1.181,0.428,0.796,0.01,0.3,14.93,1.5,0.5,30.0,0.211,0.146,0.357
1.007,1.181,0.428,0.796,0.01,0.3,14.93,1.5,0.7,44.4,0.135,0.071,0.206
1.007,1.181,0.428,0.796,0.01,0.3,14.93,1.5,0.9,64.2,0.054,0.027,0.080
1.007,1.181,0.428,0.796,0.01,0.3,19.90,2.0,0.0,0.0,0.274,0.546,0.820
1.007,1.181,0.428,0.796,0.01,0.3,19.90,2.0,0.3,17.5,0.304,0.442,0.746
1.007,1.181,0.428,0.796,0.01,0.3,19.90,2.0,0.5,30.0,0.240,0.301,0.541
1.007,1.181,0.428,0.796,0.01,0.3,19.90,2.0,0.7,44.4,0.157,0.113,0.270
1.007,1.181,0.428,0.796,0.01,0.3,19.90,2.0,0.9,64.2,0.064,0.031,0.095
1.007,1.181,0.428,0.796,0.01,0.3,29.85,3.0,0.0,0.0,0.277,0.723,1.000
1.007,1.181,0.428,0.796,0.01,0.3,29.85,3.0,0.3,17.5,0.307,0.664,0.971
1.007,1.181,0.428,0.796,0.01,0.3,29.85,3.0,0.5,30.0,0.249,0.542,0.791
1.007,1.181,0.428,0.796,0.01,0.3,29.85,3.0,0.7,44.4,0.172,0.258,0.430
1.007,1.181,0.428,0.796,0.01,0.3,29.85,3.0,0.9,64.2,0.075,0.050,0.125
1.007,1.181,0.665,0.908,0.01,0.4,10.43,1.0,0.0,0.0,0.179,0.058,0.237
1.007,1.181,0.665,0.908,0.01,0.4,10.43,1.0,0.3,17.5,0.188,0.126,0.314
1.007,1.181,0.665,0.908,0.01,0.4,10.43,1.0,0.5,30.0,0.146,0.140,0.286
1.007,1.181,0.665,0.908,0.01,0.4,10.43,1.0,0.7,44.4,0.079,0.029,0.109
1.007,1.181,0.665,0.908,0.01,0.4,10.43,1.0,0.9,64.2,0.027,0.009,0.036
1.007,1.181,0.665,0.908,0.01,0.4,15.64,1.5,0.0,0.0,0.355,0.299,0.653
1.007,1.181,0.665,0.908,0.01,0.4,15.64,1.5,0.3,17.5,0.302,0.155,0.456
1.007,1.181,0.665,0.908,0.01,0.4,15.64,1.5,0.5,30.0,0.244,0.199,0.443
1.007,1.181,0.665,0.908,0.01,0.4,15.64,1.5,0.7,44.4,0.156,0.100,0.256
1.007,1.181,0.665,0.908,0.01,0.4,15.64,1.5,0.9,64.2,0.062,0.032,0.094
1.007,1.181,0.665,0.908,0.01,0.4,20.85,2.0,0.0,0.0,0.354,0.491,0.845
1.007,1.181,0.665,0.908,0.01,0.4,20.85,2.0,0.3,17.5,0.345,0.495,0.840
1.007,1.181,0.665,0.908,0.01,0.4,20.85,2.0,0.5,30.0,0.268,0.396,0.664
1.007,1.181,0.665,0.908,0.01,0.4,20.85,2.0,0.7,44.4,0.176,0.146,0.321
1.007,1.181,0.665,0.908,0.01,0.4,20.85,2.0,0.9,64.2,0.073,0.037,0.109
1.007,1.181,0.665,0.908,0.01,0.4,31.28,3.0,0.0,0.0,0.358,0.642,1.000
1.007,1.181,0.665,0.908,0.01,0.4,31.28,3.0,0.3,17.5,0.348,0.646,0.994
1.007,1.181,0.665,0.908,0.01,0.4,31.28,3.0,0.5,30.0,0.273,0.603,0.876
1.007,1.181,0.665,0.908,0.01,0.4,31.28,3.0,0.7,44.4,0.187,0.332,0.519
1.007,1.181,0.665,0.908,0.01,0.4,31.28,3.0,0.9,64.2,0.083,0.058,0.140
1.007,1.181,0.997,1.019,0.01,0.5,11.10,1.0,0.0,0.0,0.292,0.181,0.473
1.007,1.181,0.997,1.019,0.01,0.5,11.10,1.0,0.3,17.5,0.188,0.092,0.280
1.007,1.181,0.997,1.019,0.01,0.5,11.10,1.0,0.5,30.0,0.180,0.125,0.305
1.007,1.181,0.997,1.019,0.01,0.5,11.10,1.0,0.7,44.4,0.104,0.047,0.150
1.007,1.181,0.997,1.019,0.01,0.5,11.10,1.0,0.9,64.2,0.028,0.008,0.036
1.007,1.181,0.997,1.019,0.01,0.5,16.64,1.5,0.0,0.0,0.469,0.135,0.604
1.007,1.181,0.997,1.019,0.01,0.5,16.64,1.5,0.3,17.5,0.322,0.155,0.477
1.007,1.181,0.997,1.019,0.01,0.5,16.64,1.5,0.5,30.0,0.278,0.258,0.536
1.007,1.181,0.997,1.019,0.01,0.5,16.64,1.5,0.7,44.4,0.177,0.124,0.302
1.007,1.181,0.997,1.019,0.01,0.5,16.64,1.5,0.9,64.2,0.073,0.040,0.113
1.007,1.181,0.997,1.019,0.01,0.5,22.19,2.0,0.0,0.0,0.482,0.353,0.835
1.007,1.181,0.997,1.019,0.01,0.5,22.19,2.0,0.3,17.5,0.382,0.527,0.909
1.007,1.181,0.997,1.019,0.01,0.5,22.19,2.0,0.5,30.0,0.293,0.453,0.745
1.007,1.181,0.997,1.019,0.01,0.5,22.19,2.0,0.7,44.4,0.194,0.199,0.394
1.007,1.181,0.997,1.019,0.01,0.5,22.19,2.0,0.9,64.2,0.082,0.042,0.124
1.007,1.181,0.997,1.019,0.01,0.5,33.29,3.0,0.0,0.0,0.486,0.514,1.000
1.007,1.181,0.997,1.019,0.01,0.5,33.29,3.0,0.3,17.5,0.385,0.613,0.998
1.007,1.181,0.997,1.019,0.01,0.5,33.29,3.0,0.5,30.0,0.295,0.701,0.996
1.007,1.181,0.997,1.019,0.01,0.5,33.29,3.0,0.7,44.4,0.203,0.406,0.609
1.007,1.181,0.997,1.019,0.01,0.5,33.29,3.0,0.9,64.2,0.090,0.065,0.155
,,,,,,,,,,,,
1.050,1.375,0.111,0.518,0.05,0.1,9.76,1.0,0.0,0.0,0.005,0.000,0.005
1.050,1.375,0.111,0.518,0.05,0.1,9.76,1.0,0.3,17.5,0.020,0.000,0.020
1.050,1.375,0.111,0.518,0.05,0.1,9.76,1.0,0.5,30.0,0.027,0.000,0.027
1.050,1.375,0.111,0.518,0.05,0.1,9.76,1.0,0.7,44.4,0.029,0.002,0.031
1.050,1.375,0.111,0.518,0.05,0.1,9.76,1.0,0.9,64.2,0.018,0.004,0.022
1.050,1.375,0.111,0.518,0.05,0.1,14.63,1.5,0.0,0.0,0.117,0.068,0.185
1.050,1.375,0.111,0.518,0.05,0.1,14.63,1.5,0.3,17.5,0.117,0.034,0.151
1.050,1.375,0.111,0.518,0.05,0.1,14.63,1.5,0.5,30.0,0.117,0.013,0.130
1.050,1.375,0.111,0.518,0.05,0.1,14.63,1.5,0.7,44.4,0.091,0.017,0.109
1.050,1.375,0.111,0.518,0.05,0.1,14.63,1.5,0.9,64.2,0.049,0.019,0.067
1.050,1.375,0.111,0.518,0.05,0.1,19.51,2.0,0.0,0.0,0.209,0.219,0.428
1.050,1.375,0.111,0.518,0.05,0.1,19.51,2.0,0.3,17.5,0.178,0.159,0.337
1.050,1.375,0.111,0.518,0.05,0.1,19.51,2.0,0.5,30.0,0.179,0.071,0.250
1.050,1.375,0.111,0.518,0.05,0.1,19.51,2.0,0.7,44.4,0.124,0.025,0.149
1.050,1.375,0.111,0.518,0.05,0.1,19.51,2.0,0.9,64.2,0.064,0.019,0.083
1.050,1.375,0.111,0.518,0.05,0.1,29.27,3.0,0.0,0.0,0.212,0.609,0.821
1.050,1.375,0.111,0.518,0.05,0.1,29.27,3.0,0.3,17.5,0.224,0.432,0.656
1.050,1.375,0.111,0.518,0.05,0.1,29.27,3.0,0.5,30.0,0.231,0.223,0.454
1.050,1.375,0.111,0.518,0.05,0.1,29.27,3.0,0.7,44.4,0.166,0.069,0.235
1.050,1.375,0.111,0.518,0.05,0.1,29.27,3.0,0.9,64.2,0.085,0.034,0.120
1.050,1.375,0.249,0.670,0.05,0.2,9.85,1.0,0.0,0.0,0.025,0.003,0.028
1.050,1.375,0.249,0.670,0.05,0.2,9.85,1.0,0.3,17.5,0.050,0.004,0.053
1.050,1.375,0.249,0.670,0.05,0.2,9.85,1.0,0.5,30.0,0.037,0.000,0.037
1.050,1.375,0.249,0.670,0.05,0.2,9.85,1.0,0.7,44.4,0.038,0.005,0.042
1.050,1.375,0.249,0.670,0.05,0.2,9.85,1.0,0.9,64.2,0.017,0.001,0.018
1.050,1.375,0.249,0.670,0.05,0.2,14.77,1.5,0.0,0.0,0.262,0.099,0.361
1.050,1.375,0.249,0.670,0.05,0.2,14.77,1.5,0.3,17.5,0.219,0.033,0.252
1.050,1.375,0.249,0.670,0.05,0.2,14.77,1.5,0.5,30.0,0.169,0.016,0.185
1.050,1.375,0.249,0.670,0.05,0.2,14.77,1.5,0.7,44.4,0.116,0.026,0.141
1.050,1.375,0.249,0.670,0.05,0.2,14.77,1.5,0.9,64.2,0.055,0.021,0.076
1.050,1.375,0.249,0.670,0.05,0.2,19.69,2.0,0.0,0.0,0.283,0.385,0.668
1.050,1.375,0.249,0.670,0.05,0.2,19.69,2.0,0.3,17.5,0.316,0.204,0.520
1.050,1.375,0.249,0.670,0.05,0.2,19.69,2.0,0.5,30.0,0.241,0.117,0.357
1.050,1.375,0.249,0.670,0.05,0.2,19.69,2.0,0.7,44.4,0.153,0.037,0.190
1.050,1.375,0.249,0.670,0.05,0.2,19.69,2.0,0.9,64.2,0.074,0.023,0.096
1.050,1.375,0.249,0.670,0.05,0.2,29.54,3.0,0.0,0.0,0.285,0.679,0.963
1.050,1.375,0.249,0.670,0.05,0.2,29.54,3.0,0.3,17.5,0.341,0.495,0.836
1.050,1.375,0.249,0.670,0.05,0.2,29.54,3.0,0.5,30.0,0.289,0.312,0.601
1.050,1.375,0.249,0.670,0.05,0.2,29.54,3.0,0.7,44.4,0.198,0.105,0.303
1.050,1.375,0.249,0.670,0.05,0.2,29.54,3.0,0.9,64.2,0.095,0.041,0.136
1.050,1.375,0.428,0.790,0.05,0.3,10.14,1.0,0.0,0.0,0.115,0.020,0.136
1.050,1.375,0.428,0.790,0.05,0.3,10.14,1.0,0.3,17.5,0.071,0.002,0.074
1.050,1.375,0.428,0.790,0.05,0.3,10.14,1.0,0.5,30.0,0.052,0.002,0.054
1.050,1.375,0.428,0.790,0.05,0.3,10.14,1.0,0.7,44.4,0.041,0.003,0.044
1.050,1.375,0.428,0.790,0.05,0.3,10.14,1.0,0.9,64.2,0.014,0.001,0.015
1.050,1.375,0.428,0.790,0.05,0.3,15.21,1.5,0.0,0.0,0.339,0.105,0.444
1.050,1.375,0.428,0.790,0.05,0.3,15.21,1.5,0.3,17.5,0.283,0.040,0.324
1.050,1.375,0.428,0.790,0.05,0.3,15.21,1.5,0.5,30.0,0.214,0.064,0.278
1.050,1.375,0.428,0.790,0.05,0.3,15.21,1.5,0.7,44.4,0.136,0.033,0.170
1.050,1.375,0.428,0.790,0.05,0.3,15.21,1.5,0.9,64.2,0.061,0.022,0.083
1.050,1.375,0.428,0.790,0.05,0.3,20.28,2.0,0.0,0.0,0.351,0.401,0.752
1.050,1.375,0.428,0.790,0.05,0.3,20.28,2.0,0.3,17.5,0.386,0.276,0.662
1.050,1.375,0.428,0.790,0.05,0.3,20.28,2.0,0.5,30.0,0.288,0.156,0.445
1.050,1.375,0.428,0.790,0.05,0.3,20.28,2.0,0.7,44.4,0.179,0.050,0.229
1.050,1.375,0.428,0.790,0.05,0.3,20.28,2.0,0.9,64.2,0.082,0.025,0.107
1.050,1.375,0.428,0.790,0.05,0.3,30.42,3.0,0.0,0.0,0.357,0.642,1.000
1.050,1.375,0.428,0.790,0.05,0.3,30.42,3.0,0.3,17.5,0.396,0.536,0.932
1.050,1.375,0.428,0.790,0.05,0.3,30.42,3.0,0.5,30.0,0.332,0.380,0.712
1.050,1.375,0.428,0.790,0.05,0.3,30.42,3.0,0.7,44.4,0.224,0.145,0.369
1.050,1.375,0.428,0.790,0.05,0.3,30.42,3.0,0.9,64.2,0.104,0.046,0.150
1.050,1.375,0.665,0.902,0.05,0.4,10.60,1.0,0.0,0.0,0.159,0.050,0.208
1.050,1.375,0.665,0.902,0.05,0.4,10.60,1.0,0.3,17.5,0.104,0.010,0.114
1.050,1.375,0.665,0.902,0.05,0.4,10.60,1.0,0.5,30.0,0.072,0.017,0.089
1.050,1.375,0.665,0.902,0.05,0.4,10.60,1.0,0.7,44.4,0.042,0.002,0.044
1.050,1.375,0.665,0.902,0.05,0.4,10.60,1.0,0.9,64.2,0.014,0.000,0.014
1.050,1.375,0.665,0.902,0.05,0.4,15.90,1.5,0.0,0.0,0.409,0.069,0.478
1.050,1.375,0.665,0.902,0.05,0.4,15.90,1.5,0.3,17.5,0.338,0.056,0.394
1.050,1.375,0.665,0.902,0.05,0.4,15.90,1.5,0.5,30.0,0.264,0.088,0.352
1.050,1.375,0.665,0.902,0.05,0.4,15.90,1.5,0.7,44.4,0.160,0.044,0.204
1.050,1.375,0.665,0.902,0.05,0.4,15.90,1.5,0.9,64.2,0.070,0.024,0.094
1.050,1.375,0.665,0.902,0.05,0.4,21.20,2.0,0.0,0.0,0.433,0.375,0.808
1.050,1.375,0.665,0.902,0.05,0.4,21.20,2.0,0.3,17.5,0.432,0.336,0.768
1.050,1.375,0.665,0.902,0.05,0.4,21.20,2.0,0.5,30.0,0.337,0.202,0.538
1.050,1.375,0.665,0.902,0.05,0.4,21.20,2.0,0.7,44.4,0.204,0.067,0.271
1.050,1.375,0.665,0.902,0.05,0.4,21.20,2.0,0.9,64.2,0.090,0.028,0.118
1.050,1.375,0.665,0.902,0.05,0.4,31.81,3.0,0.0,0.0,0.440,0.560,1.000
1.050,1.375,0.665,0.902,0.05,0.4,31.81,3.0,0.3,17.5,0.439,0.538,0.977
1.050,1.375,0.665,0.902,0.05,0.4,31.81,3.0,0.5,30.0,0.361,0.447,0.808
1.050,1.375,0.665,0.902,0.05,0.4,31.81,3.0,0.7,44.4,0.248,0.190,0.437
1.050,1.375,0.665,0.902,0.05,0.4,31.81,3.0,0.9,64.2,0.113,0.052,0.165
1.050,1.375,0.997,1.013,0.05,0.5,11.26,1.0,0.0,0.0,0.199,0.009,0.208
1.050,1.375,0.997,1.013,0.05,0.5,11.26,1.0,0.3,17.5,0.146,0.020,0.165
1.050,1.375,0.997,1.013,0.05,0.5,11.26,1.0,0.5,30.0,0.096,0.018,0.114
1.050,1.375,0.997,1.013,0.05,0.5,11.26,1.0,0.7,44.4,0.049,0.003,0.052
1.050,1.375,0.997,1.013,0.05,0.5,11.26,1.0,0.9,64.2,0.014,0.000,0.015
1.050,1.375,0.997,1.013,0.05,0.5,16.89,1.5,0.0,0.0,0.451,0.033,0.484
1.050,1.375,0.997,1.013,0.05,0.5,16.89,1.5,0.3,17.5,0.368,0.081,0.449
1.050,1.375,0.997,1.013,0.05,0.5,16.89,1.5,0.5,30.0,0.317,0.120,0.437
1.050,1.375,0.997,1.013,0.05,0.5,16.89,1.5,0.7,44.4,0.189,0.060,0.249
1.050,1.375,0.997,1.013,0.05,0.5,16.89,1.5,0.9,64.2,0.079,0.026,0.105
1.050,1.375,0.997,1.013,0.05,0.5,22.52,2.0,0.0,0.0,0.549,0.264,0.812
1.050,1.375,0.997,1.013,0.05,0.5,22.52,2.0,0.3,17.5,0.472,0.386,0.857
1.050,1.375,0.997,1.013,0.05,0.5,22.52,2.0,0.5,30.0,0.376,0.273,0.649
1.050,1.375,0.997,1.013,0.05,0.5,22.52,2.0,0.7,44.4,0.233,0.092,0.325
1.050,1.375,0.997,1.013,0.05,0.5,22.52,2.0,0.9,64.2,0.101,0.032,0.133
1.050,1.375,0.997,1.013,0.05,0.5,33.77,3.0,0.0,0.0,0.558,0.442,1.000
1.050,1.375,0.997,1.013,0.05,0.5,33.77,3.0,0.3,17.5,0.478,0.519,0.997
1.050,1.375,0.997,1.013,0.05,0.5,33.77,3.0,0.5,30.0,0.387,0.512,0.899
1.050,1.375,0.997,1.013,0.05,0.5,33.77,3.0,0.7,44.4,0.270,0.253,0.523
1.050,1.375,0.997,1.013,0.05,0.5,33.77,3.0,0.9,64.2,0.122,0.061,0.183
,,,,,,,,,,,,
1.108,1.562,0.111,0.518,0.1,0.1,10.01,1.0,0.0,0.0,0.001,0.000,0.001
1.108,1.562,0.111,0.518,0.1,0.1,10.01,1.0,0.3,17.5,0.006,0.000,0.006
1.108,1.562,0.111,0.518,0.1,0.1,10.01,1.0,0.5,30.0,0.013,0.000,0.013
1.108,1.562,0.111,0.518,0.1,0.1,10.01,1.0,0.7,44.4,0.019,0.001,0.021
1.108,1.562,0.111,0.518,0.1,0.1,10.01,1.0,0.9,64.2,0.017,0.001,0.017
1.108,1.562,0.111,0.518,0.1,0.1,15.02,1.5,0.0,0.0,0.061,0.022,0.083
1.108,1.562,0.111,0.518,0.1,0.1,15.02,1.5,0.3,17.5,0.081,0.017,0.098
1.108,1.562,0.111,0.518,0.1,0.1,15.02,1.5,0.5,30.0,0.094,0.004,0.099
1.108,1.562,0.111,0.518,0.1,0.1,15.02,1.5,0.7,44.4,0.092,0.012,0.104
1.108,1.562,0.111,0.518,0.1,0.1,15.02,1.5,0.9,64.2,0.058,0.017,0.076
1.108,1.562,0.111,0.518,0.1,0.1,20.02,2.0,0.0,0.0,0.212,0.111,0.323
1.108,1.562,0.111,0.518,0.1,0.1,20.02,2.0,0.3,17.5,0.171,0.085,0.256
1.108,1.562,0.111,0.518,0.1,0.1,20.02,2.0,0.5,30.0,0.168,0.038,0.206
1.108,1.562,0.111,0.518,0.1,0.1,20.02,2.0,0.7,44.4,0.127,0.018,0.145
1.108,1.562,0.111,0.518,0.1,0.1,20.02,2.0,0.9,64.2,0.075,0.021,0.096
1.108,1.562,0.111,0.518,0.1,0.1,30.03,3.0,0.0,0.0,0.265,0.491,0.756
1.108,1.562,0.111,0.518,0.1,0.1,30.03,3.0,0.3,17.5,0.257,0.326,0.583
1.108,1.562,0.111,0.518,0.1,0.1,30.03,3.0,0.5,30.0,0.253,0.147,0.400
1.108,1.562,0.111,0.518,0.1,0.1,30.03,3.0,0.7,44.4,0.178,0.048,0.226
1.108,1.562,0.111,0.518,0.1,0.1,30.03,3.0,0.9,64.2,0.102,0.036,0.138
1.108,1.562,0.249,0.670,0.1,0.2,10.07,1.0,0.0,0.0,0.005,0.016,0.021
1.108,1.562,0.249,0.670,0.1,0.2,10.07,1.0,0.3,17.5,0.022,0.000,0.023
1.108,1.562,0.249,0.670,0.1,0.2,10.07,1.0,0.5,30.0,0.022,0.000,0.022
1.108,1.562,0.249,0.670,0.1,0.2,10.07,1.0,0.7,44.4,0.026,0.001,0.026
1.108,1.562,0.249,0.670,0.1,0.2,10.07,1.0,0.9,64.2,0.016,0.000,0.016
1.108,1.562,0.249,0.670,0.1,0.2,15.11,1.5,0.0,0.0,0.244,0.053,0.297
1.108,1.562,0.249,0.670,0.1,0.2,15.11,1.5,0.3,17.5,0.173,0.017,0.191
1.108,1.562,0.249,0.670,0.1,0.2,15.11,1.5,0.5,30.0,0.151,0.010,0.161
1.108,1.562,0.249,0.670,0.1,0.2,15.11,1.5,0.7,44.4,0.117,0.020,0.137
1.108,1.562,0.249,0.670,0.1,0.2,15.11,1.5,0.9,64.2,0.063,0.020,0.083
1.108,1.562,0.249,0.670,0.1,0.2,20.15,2.0,0.0,0.0,0.334,0.247,0.581
1.108,1.562,0.249,0.670,0.1,0.2,20.15,2.0,0.3,17.5,0.329,0.114,0.443
1.108,1.562,0.249,0.670,0.1,0.2,20.15,2.0,0.5,30.0,0.238,0.074,0.311
1.108,1.562,0.249,0.670,0.1,0.2,20.15,2.0,0.7,44.4,0.159,0.028,0.187
1.108,1.562,0.249,0.670,0.1,0.2,20.15,2.0,0.9,64.2,0.085,0.023,0.108
1.108,1.562,0.249,0.670,0.1,0.2,30.22,3.0,0.0,0.0,0.344,0.602,0.946
1.108,1.562,0.249,0.670,0.1,0.2,30.22,3.0,0.3,17.5,0.401,0.390,0.790
1.108,1.562,0.249,0.670,0.1,0.2,30.22,3.0,0.5,30.0,0.326,0.222,0.548
1.108,1.562,0.249,0.670,0.1,0.2,30.22,3.0,0.7,44.4,0.218,0.078,0.296
1.108,1.562,0.249,0.670,0.1,0.2,30.22,3.0,0.9,64.2,0.114,0.043,0.157
1.108,1.562,0.428,0.790,0.1,0.3,10.35,1.0,0.0,0.0,0.057,0.002,0.059
1.108,1.562,0.428,0.790,0.1,0.3,10.35,1.0,0.3,17.5,0.035,0.004,0.039
1.108,1.562,0.428,0.790,0.1,0.3,10.35,1.0,0.5,30.0,0.030,0.001,0.031
1.108,1.562,0.428,0.790,0.1,0.3,10.35,1.0,0.7,44.4,0.032,0.001,0.034
1.108,1.562,0.428,0.790,0.1,0.3,10.35,1.0,0.9,64.2,0.013,0.000,0.013
1.108,1.562,0.428,0.790,0.1,0.3,15.52,1.5,0.0,0.0,0.337,0.067,0.404
1.108,1.562,0.428,0.790,0.1,0.3,15.52,1.5,0.3,17.5,0.255,0.019,0.274
1.108,1.562,0.428,0.790,0.1,0.3,15.52,1.5,0.5,30.0,0.202,0.039,0.241
1.108,1.562,0.428,0.790,0.1,0.3,15.52,1.5,0.7,44.4,0.138,0.025,0.163
1.108,1.562,0.428,0.790,0.1,0.3,15.52,1.5,0.9,64.2,0.071,0.021,0.091
1.108,1.562,0.428,0.790,0.1,0.3,20.70,2.0,0.0,0.0,0.407,0.288,0.695
1.108,1.562,0.428,0.790,0.1,0.3,20.70,2.0,0.3,17.5,0.425,0.169,0.594
1.108,1.562,0.428,0.790,0.1,0.3,20.70,2.0,0.5,30.0,0.294,0.100,0.394
1.108,1.562,0.428,0.790,0.1,0.3,20.70,2.0,0.7,44.4,0.186,0.037,0.223
1.108,1.562,0.428,0.790,0.1,0.3,20.70,2.0,0.9,64.2,0.093,0.025,0.118
1.108,1.562,0.428,0.790,0.1,0.3,31.05,3.0,0.0,0.0,0.422,0.572,0.994
1.108,1.562,0.428,0.790,0.1,0.3,31.05,3.0,0.3,17.5,0.460,0.447,0.907
1.108,1.562,0.428,0.790,0.1,0.3,31.05,3.0,0.5,30.0,0.380,0.281,0.661
1.108,1.562,0.428,0.790,0.1,0.3,31.05,3.0,0.7,44.4,0.249,0.108,0.357
1.108,1.562,0.428,0.790,0.1,0.3,31.05,3.0,0.9,64.2,0.123,0.049,0.172
1.108,1.562,0.665,0.902,0.1,0.4,10.79,1.0,0.0,0.0,0.110,0.031,0.140
1.108,1.562,0.665,0.902,0.1,0.4,10.79,1.0,0.3,17.5,0.059,0.003,0.062
1.108,1.562,0.665,0.902,0.1,0.4,10.79,1.0,0.5,30.0,0.045,0.006,0.051
1.108,1.562,0.665,0.902,0.1,0.4,10.79,1.0,0.7,44.4,0.032,0.001,0.033
1.108,1.562,0.665,0.902,0.1,0.4,10.79,1.0,0.9,64.2,0.013,0.000,0.013
1.108,1.562,0.665,0.902,0.1,0.4,16.19,1.5,0.0,0.0,0.395,0.040,0.436
1.108,1.562,0.665,0.902,0.1,0.4,16.19,1.5,0.3,17.5,0.342,0.028,0.370
1.108,1.562,0.665,0.902,0.1,0.4,16.19,1.5,0.5,30.0,0.256,0.056,0.312
1.108,1.562,0.665,0.902,0.1,0.4,16.19,1.5,0.7,44.4,0.162,0.032,0.194
1.108,1.562,0.665,0.902,0.1,0.4,16.19,1.5,0.9,64.2,0.077,0.022,0.099
1.108,1.562,0.665,0.902,0.1,0.4,21.58,2.0,0.0,0.0,0.493,0.271,0.764
1.108,1.562,0.665,0.902,0.1,0.4,21.58,2.0,0.3,17.5,0.489,0.222,0.711
1.108,1.562,0.665,0.902,0.1,0.4,21.58,2.0,0.5,30.0,0.354,0.130,0.483
1.108,1.562,0.665,0.902,0.1,0.4,21.58,2.0,0.7,44.4,0.214,0.049,0.263
1.108,1.562,0.665,0.902,0.1,0.4,21.58,2.0,0.9,64.2,0.103,0.028,0.131
1.108,1.562,0.665,0.902,0.1,0.4,32.37,3.0,0.0,0.0,0.507,0.493,1.000
1.108,1.562,0.665,0.902,0.1,0.4,32.37,3.0,0.3,17.5,0.505,0.463,0.968
1.108,1.562,0.665,0.902,0.1,0.4,32.37,3.0,0.5,30.0,0.422,0.343,0.765
1.108,1.562,0.665,0.902,0.1,0.4,32.37,3.0,0.7,44.4,0.278,0.142,0.420
1.108,1.562,0.665,0.902,0.1,0.4,32.37,3.0,0.9,64.2,0.133,0.055,0.189
1.108,1.562,0.997,1.013,0.1,0.5,11.43,1.0,0.0,0.0,0.179,0.006,0.185
1.108,1.562,0.997,1.013,0.1,0.5,11.43,1.0,0.3,17.5,0.092,0.010,0.101
1.108,1.562,0.997,1.013,0.1,0.5,11.43,1.0,0.5,30.0,0.075,0.009,0.084
1.108,1.562,0.997,1.013,0.1,0.5,11.43,1.0,0.7,44.4,0.037,0.001,0.037
1.108,1.562,0.997,1.013,0.1,0.5,11.43,1.0,0.9,64.2,0.012,0.000,0.012
1.108,1.562,0.997,1.013,0.1,0.5,17.14,1.5,0.0,0.0,0.434,0.016,0.449
1.108,1.562,0.997,1.013,0.1,0.5,17.14,1.5,0.3,17.5,0.403,0.057,0.460
1.108,1.562,0.997,1.013,0.1,0.5,17.14,1.5,0.5,30.0,0.325,0.079,0.403
1.108,1.562,0.997,1.013,0.1,0.5,17.14,1.5,0.7,44.4,0.190,0.040,0.230
1.108,1.562,0.997,1.013,0.1,0.5,17.14,1.5,0.9,64.2,0.086,0.023,0.108
1.108,1.562,0.997,1.013,0.1,0.5,22.85,2.0,0.0,0.0,0.612,0.164,0.777
1.108,1.562,0.997,1.013,0.1,0.5,22.85,2.0,0.3,17.5,0.532,0.283,0.814
1.108,1.562,0.997,1.013,0.1,0.5,22.85,2.0,0.5,30.0,0.416,0.170,0.586
1.108,1.562,0.997,1.013,0.1,0.5,22.85,2.0,0.7,44.4,0.247,0.066,0.313
1.108,1.562,0.997,1.013,0.1,0.5,22.85,2.0,0.9,64.2,0.113,0.030,0.143
1.108,1.562,0.997,1.013,0.1,0.5,34.28,3.0,0.0,0.0,0.628,0.372,1.000
1.108,1.562,0.997,1.013,0.1,0.5,34.28,3.0,0.3,17.5,0.545,0.447,0.992
1.108,1.562,0.997,1.013,0.1,0.5,34.28,3.0,0.5,30.0,0.454,0.411,0.865
1.108,1.562,0.997,1.013,0.1,0.5,34.28,3.0,0.7,44.4,0.309,0.186,0.495
1.108,1.562,0.997,1.013,0.1,0.5,34.28,3.0,0.9,64.2,0.145,0.065,0.210
,,,,,,,,,,,,
1.247,1.941,0.111,0.518,0.2,0.1,10.59,1.0,0.0,0.0,0.001,0.000,0.001
1.247,1.941,0.111,0.518,0.2,0.1,10.59,1.0,0.3,17.5,0.002,0.000,0.002
1.247,1.941,0.111,0.518,0.2,0.1,10.59,1.0,0.5,30.0,0.006,0.000,0.006
1.247,1.941,0.111,0.518,0.2,0.1,10.59,1.0,0.7,44.4,0.010,0.001,0.011
1.247,1.941,0.111,0.518,0.2,0.1,10.59,1.0,0.9,64.2,0.014,0.002,0.016
1.247,1.941,0.111,0.518,0.2,0.1,15.88,1.5,0.0,0.0,0.013,0.006,0.019
1.247,1.941,0.111,0.518,0.2,0.1,15.88,1.5,0.3,17.5,0.047,0.007,0.054
1.247,1.941,0.111,0.518,0.2,0.1,15.88,1.5,0.5,30.0,0.072,0.002,0.074
1.247,1.941,0.111,0.518,0.2,0.1,15.88,1.5,0.7,44.4,0.081,0.009,0.090
1.247,1.941,0.111,0.518,0.2,0.1,15.88,1.5,0.9,64.2,0.066,0.014,0.080
1.247,1.941,0.111,0.518,0.2,0.1,21.17,2.0,0.0,0.0,0.136,0.052,0.188
1.247,1.941,0.111,0.518,0.2,0.1,21.17,2.0,0.3,17.5,0.127,0.039,0.166
1.247,1.941,0.111,0.518,0.2,0.1,21.17,2.0,0.5,30.0,0.147,0.017,0.164
1.247,1.941,0.111,0.518,0.2,0.1,21.17,2.0,0.7,44.4,0.127,0.016,0.143
1.247,1.941,0.111,0.518,0.2,0.1,21.17,2.0,0.9,64.2,0.090,0.020,0.110
1.247,1.941,0.111,0.518,0.2,0.1,31.76,3.0,0.0,0.0,0.315,0.297,0.612
1.247,1.941,0.111,0.518,0.2,0.1,31.76,3.0,0.3,17.5,0.274,0.212,0.486
1.247,1.941,0.111,0.518,0.2,0.1,31.76,3.0,0.5,30.0,0.257,0.087,0.344
1.247,1.941,0.111,0.518,0.2,0.1,31.76,3.0,0.7,44.4,0.190,0.038,0.228
1.247,1.941,0.111,0.518,0.2,0.1,31.76,3.0,0.9,64.2,0.124,0.037,0.161
1.247,1.941,0.249,0.670,0.2,0.2,10.60,1.0,0.0,0.0,0.001,0.014,0.015
1.247,1.941,0.249,0.670,0.2,0.2,10.60,1.0,0.3,17.5,0.007,0.000,0.007
1.247,1.941,0.249,0.670,0.2,0.2,10.60,1.0,0.5,30.0,0.010,0.000,0.010
1.247,1.941,0.249,0.670,0.2,0.2,10.60,1.0,0.7,44.4,0.016,0.000,0.016
1.247,1.941,0.249,0.670,0.2,0.2,10.60,1.0,0.9,64.2,0.017,0.002,0.019
1.247,1.941,0.249,0.670,0.2,0.2,15.89,1.5,0.0,0.0,0.144,0.023,0.167
1.247,1.941,0.249,0.670,0.2,0.2,15.89,1.5,0.3,17.5,0.118,0.009,0.127
1.247,1.941,0.249,0.670,0.2,0.2,15.89,1.5,0.5,30.0,0.128,0.007,0.134
1.247,1.941,0.249,0.670,0.2,0.2,15.89,1.5,0.7,44.4,0.115,0.015,0.130
1.247,1.941,0.249,0.670,0.2,0.2,15.89,1.5,0.9,64.2,0.078,0.017,0.095
1.247,1.941,0.249,0.670,0.2,0.2,21.19,2.0,0.0,0.0,0.340,0.120,0.460
1.247,1.941,0.249,0.670,0.2,0.2,21.19,2.0,0.3,17.5,0.287,0.056,0.343
1.247,1.941,0.249,0.670,0.2,0.2,21.19,2.0,0.5,30.0,0.221,0.043,0.264
1.247,1.941,0.249,0.670,0.2,0.2,21.19,2.0,0.7,44.4,0.165,0.025,0.190
1.247,1.941,0.249,0.670,0.2,0.2,21.19,2.0,0.9,64.2,0.102,0.025,0.127
1.247,1.941,0.249,0.670,0.2,0.2,31.79,3.0,0.0,0.0,0.410,0.505,0.915
1.247,1.941,0.249,0.670,0.2,0.2,31.79,3.0,0.3,17.5,0.453,0.265,0.717
1.247,1.941,0.249,0.670,0.2,0.2,31.79,3.0,0.5,30.0,0.345,0.142,0.486
1.247,1.941,0.249,0.670,0.2,0.2,31.79,3.0,0.7,44.4,0.236,0.063,0.299
1.247,1.941,0.249,0.670,0.2,0.2,31.79,3.0,0.9,64.2,0.140,0.048,0.188
1.247,1.941,0.428,0.790,0.2,0.3,10.83,1.0,0.0,0.0,0.006,0.000,0.006
1.247,1.941,0.428,0.790,0.2,0.3,10.83,1.0,0.3,17.5,0.013,0.000,0.013
1.247,1.941,0.428,0.790,0.2,0.3,10.83,1.0,0.5,30.0,0.016,0.000,0.016
1.247,1.941,0.428,0.790,0.2,0.3,10.83,1.0,0.7,44.4,0.023,0.001,0.024
1.247,1.941,0.428,0.790,0.2,0.3,10.83,1.0,0.9,64.2,0.017,0.003,0.019
1.247,1.941,0.428,0.790,0.2,0.3,16.24,1.5,0.0,0.0,0.262,0.033,0.295
1.247,1.941,0.428,0.790,0.2,0.3,16.24,1.5,0.3,17.5,0.195,0.011,0.206
1.247,1.941,0.428,0.790,0.2,0.3,16.24,1.5,0.5,30.0,0.181,0.023,0.205
1.247,1.941,0.428,0.790,0.2,0.3,16.24,1.5,0.7,44.4,0.140,0.020,0.160
1.247,1.941,0.428,0.790,0.2,0.3,16.24,1.5,0.9,64.2,0.084,0.019,0.103
1.247,1.941,0.428,0.790,0.2,0.3,21.65,2.0,0.0,0.0,0.454,0.155,0.609
1.247,1.941,0.428,0.790,0.2,0.3,21.65,2.0,0.3,17.5,0.422,0.095,0.517
1.247,1.941,0.428,0.790,0.2,0.3,21.65,2.0,0.5,30.0,0.283,0.062,0.346
1.247,1.941,0.428,0.790,0.2,0.3,21.65,2.0,0.7,44.4,0.195,0.034,0.229
1.247,1.941,0.428,0.790,0.2,0.3,21.65,2.0,0.9,64.2,0.112,0.028,0.139
1.247,1.941,0.428,0.790,0.2,0.3,32.48,3.0,0.0,0.0,0.493,0.483,0.977
1.247,1.941,0.428,0.790,0.2,0.3,32.48,3.0,0.3,17.5,0.529,0.329,0.859
1.247,1.941,0.428,0.790,0.2,0.3,32.48,3.0,0.5,30.0,0.415,0.183,0.598
1.247,1.941,0.428,0.790,0.2,0.3,32.48,3.0,0.7,44.4,0.272,0.086,0.358
1.247,1.941,0.428,0.790,0.2,0.3,32.48,3.0,0.9,64.2,0.152,0.057,0.209
1.247,1.941,0.665,0.902,0.2,0.4,11.22,1.0,0.0,0.0,0.028,0.000,0.028
1.247,1.941,0.665,0.902,0.2,0.4,11.22,1.0,0.3,17.5,0.024,0.001,0.026
1.247,1.941,0.665,0.902,0.2,0.4,11.22,1.0,0.5,30.0,0.025,0.002,0.027
1.247,1.941,0.665,0.902,0.2,0.4,11.22,1.0,0.7,44.4,0.027,0.002,0.029
1.247,1.941,0.665,0.902,0.2,0.4,11.22,1.0,0.9,64.2,0.014,0.001,0.016
1.247,1.941,0.665,0.902,0.2,0.4,16.83,1.5,0.0,0.0,0.347,0.031,0.378
1.247,1.941,0.665,0.902,0.2,0.4,16.83,1.5,0.3,17.5,0.307,0.019,0.325
1.247,1.941,0.665,0.902,0.2,0.4,16.83,1.5,0.5,30.0,0.241,0.035,0.276
1.247,1.941,0.665,0.902,0.2,0.4,16.83,1.5,0.7,44.4,0.163,0.025,0.187
1.247,1.941,0.665,0.902,0.2,0.4,16.83,1.5,0.9,64.2,0.091,0.021,0.112
1.247,1.941,0.665,0.902,0.2,0.4,22.44,2.0,0.0,0.0,0.553,0.146,0.698
1.247,1.941,0.665,0.902,0.2,0.4,22.44,2.0,0.3,17.5,0.526,0.120,0.645
1.247,1.941,0.665,0.902,0.2,0.4,22.44,2.0,0.5,30.0,0.350,0.080,0.430
1.247,1.941,0.665,0.902,0.2,0.4,22.44,2.0,0.7,44.4,0.225,0.042,0.267
1.247,1.941,0.665,0.902,0.2,0.4,22.44,2.0,0.9,64.2,0.122,0.031,0.153
1.247,1.941,0.665,0.902,0.2,0.4,33.67,3.0,0.0,0.0,0.586,0.414,1.000
1.247,1.941,0.665,0.902,0.2,0.4,33.67,3.0,0.3,17.5,0.578,0.368,0.946
1.247,1.941,0.665,0.902,0.2,0.4,33.67,3.0,0.5,30.0,0.476,0.230,0.706
1.247,1.941,0.665,0.902,0.2,0.4,33.67,3.0,0.7,44.4,0.306,0.111,0.417
1.247,1.941,0.665,0.902,0.2,0.4,33.67,3.0,0.9,64.2,0.164,0.067,0.231
1.247,1.941,0.997,1.013,0.2,0.5,11.82,1.0,0.0,0.0,0.072,0.002,0.074
1.247,1.941,0.997,1.013,0.2,0.5,11.82,1.0,0.3,17.5,0.044,0.003,0.047
1.247,1.941,0.997,1.013,0.2,0.5,11.82,1.0,0.5,30.0,0.048,0.005,0.053
1.247,1.941,0.997,1.013,0.2,0.5,11.82,1.0,0.7,44.4,0.029,0.000,0.029
1.247,1.941,0.997,1.013,0.2,0.5,11.82,1.0,0.9,64.2,0.014,0.000,0.015
1.247,1.941,0.997,1.013,0.2,0.5,17.72,1.5,0.0,0.0,0.407,0.016,0.423
1.247,1.941,0.997,1.013,0.2,0.5,17.72,1.5,0.3,17.5,0.412,0.020,0.431
1.247,1.941,0.997,1.013,0.2,0.5,17.72,1.5,0.5,30.0,0.313,0.049,0.362
1.247,1.941,0.997,1.013,0.2,0.5,17.72,1.5,0.7,44.4,0.192,0.031,0.224
1.247,1.941,0.997,1.013,0.2,0.5,17.72,1.5,0.9,64.2,0.101,0.022,0.122
1.247,1.941,0.997,1.013,0.2,0.5,23.63,2.0,0.0,0.0,0.663,0.093,0.756
1.247,1.941,0.997,1.013,0.2,0.5,23.63,2.0,0.3,17.5,0.592,0.177,0.769
1.247,1.941,0.997,1.013,0.2,0.5,23.63,2.0,0.5,30.0,0.426,0.103,0.529
1.247,1.941,0.997,1.013,0.2,0.5,23.63,2.0,0.7,44.4,0.261,0.052,0.313
1.247,1.941,0.997,1.013,0.2,0.5,23.63,2.0,0.9,64.2,0.132,0.036,0.167
1.247,1.941,0.997,1.013,0.2,0.5,35.45,3.0,0.0,0.0,0.708,0.292,1.000
1.247,1.941,0.997,1.013,0.2,0.5,35.45,3.0,0.3,17.5,0.621,0.362,0.983
1.247,1.941,0.997,1.013,0.2,0.5,35.45,3.0,0.5,30.0,0.526,0.288,0.814
1.247,1.941,0.997,1.013,0.2,0.5,35.45,3.0,0.7,44.4,0.341,0.141,0.482
1.247,1.941,0.997,1.013,0.2,0.5,35.45,3.0,0.9,64.2,0.176,0.079,0.255
\end{filecontents*}
\csvreader[
        head to column names, 
        late after line=\\, 
        late after last line=\\ 
    ]{impact_scenarios.csv}{}
    {%
        \csvcoli & \csvcolii & \csvcoliii & \csvcoliv & \csvcolv & \csvcolvi & \csvcolvii & \csvcolviii & \csvcolix & \csvcolx & \csvcolxi & \csvcolxii & \csvcolxiii
    }
    \label{table:initial_conditions}
\end{longtable*}
\clearpage

\section{Near- and far-field losses}\label{sec:NF_loss}

The deconvolved near- and far-field loss for the cases shown in Figure~\ref{fig:loss_data} are shown in Figures~\ref{fig:X_NF} and \ref{fig:X_FF}, respectively.

\setcounter{figure}{0}
\begin{figure*}[h]
\begin{center}
    \includegraphics[scale=0.5]{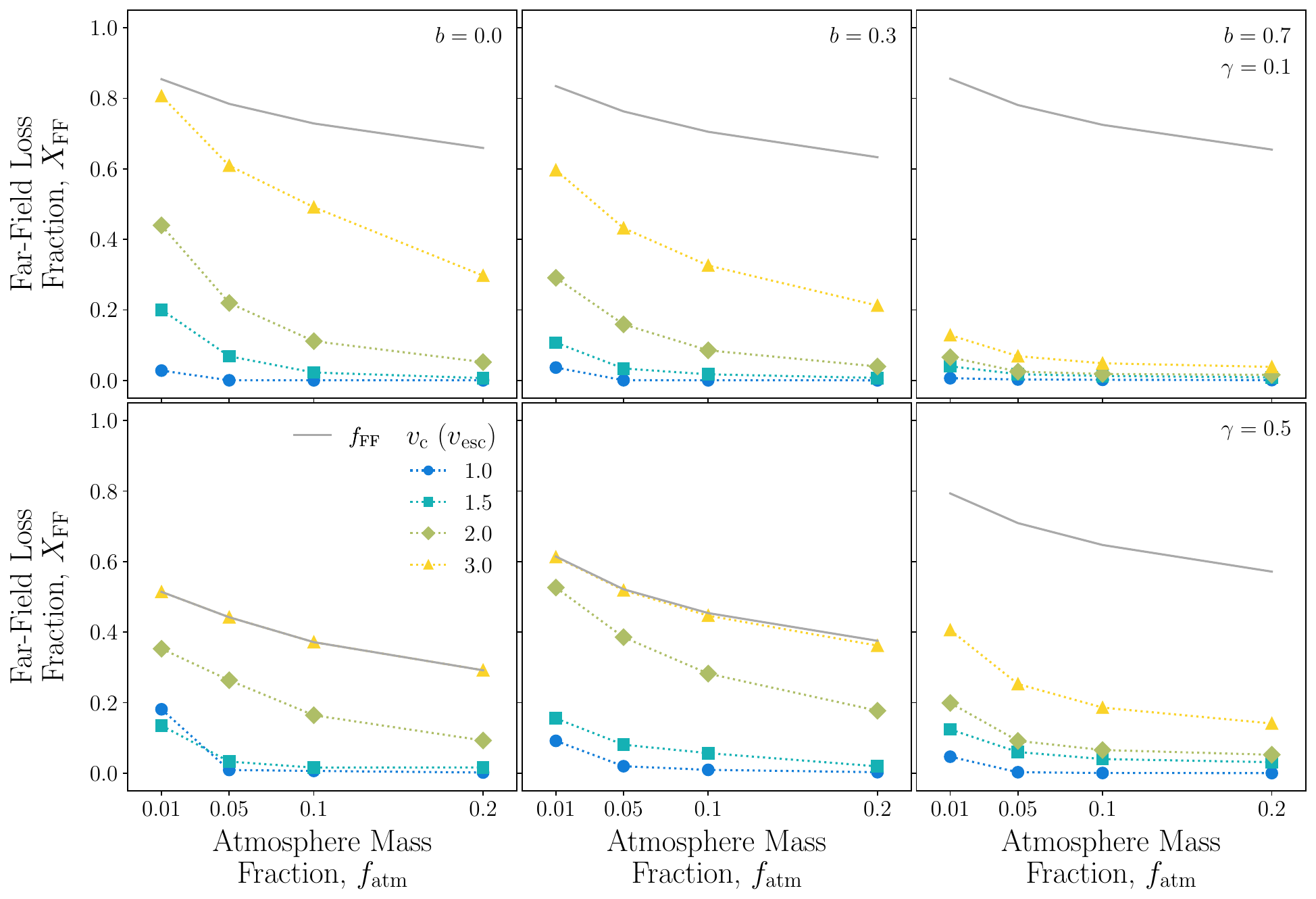}
    \caption{The same as Figure \ref{fig:loss_data} but showing atmospheric loss from the far field only. The grey solid line shows the mass of atmosphere that is in the far field.}\label{fig:X_FF}
\end{center}
\end{figure*}

\begin{figure*}[h]
\begin{center}
    \includegraphics[scale=0.5]{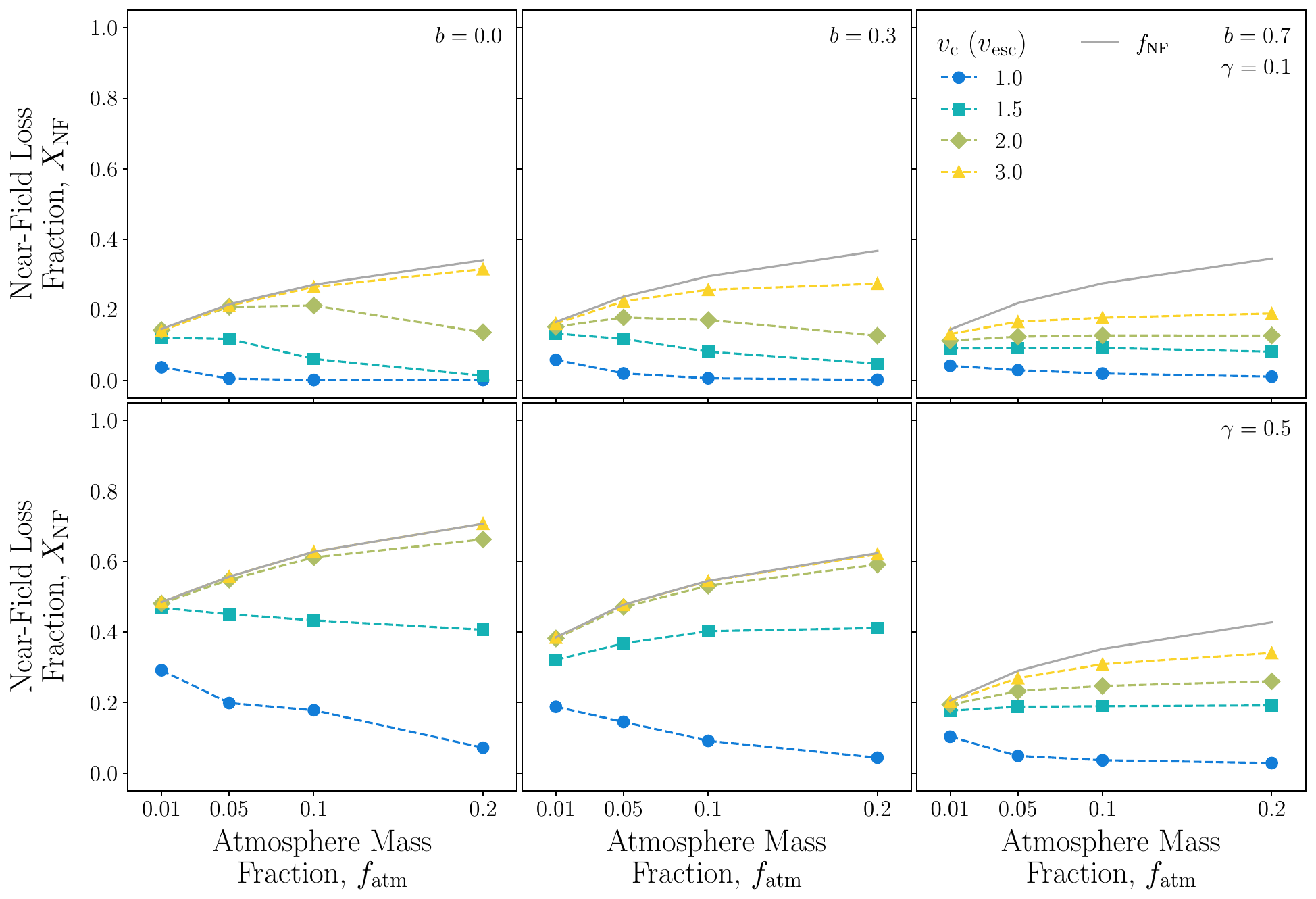}
    \caption{The same as Figure \ref{fig:loss_data} but showing atmospheric loss from the near field only. The grey solid line shows the mass of atmosphere that is in the near field.}\label{fig:X_NF}
\end{center}
\end{figure*}
\clearpage

\section{Fitting parameters}\label{sec:fitting_params}
\setcounter{table}{0}

The fitting parameters used in the scaling law for near-field mass (Eqn. \ref{eq:zeta_i}), near-field (Eqn. \ref{eq:xi_i}) and far-field (Eqn. \ref{eq:psi_i}) loss are listed below in Tables \ref{table:f_NF_params}, \ref{table:NF_params}, and \ref{table:FF_params}, respectively.

\renewcommand{\thetable}{\thesection\arabic{table}}
\begin{table}[h]
    \caption{Fitting parameters, $q_{ij}$, used in Eqn.~\ref{eq:zeta_i} to predict the near-field mass.}
    \label{table:f_NF_params}
    \centering
    \begin{tabular}{c|cccc}
    \toprule
    \toprule
         & Parameter & Value & Parameter & Value \\
         \midrule
         \multirow{4}{*}{$\zeta_1$} & $q_{11}$ & --0.00114 & $q_{15}$ & --0.000168 \\
                             & $q_{12}$ & 0.000585 & $q_{16}$ & 0.000 \\
                             & $q_{13}$ & 0.000423 & $q_{17}$ & 0.000 \\
                             & $q_{14}$ & 2.000 & $q_{18}$ & 0.000 \\
                                     
         \midrule
         \multirow{4}{*}{$\zeta_2$} & $q_{21}$ & --2.990 & $q_{25}$ & 0.000 \\
                             & $q_{22}$ & --4.341 & $q_{26}$ & 0.000 \\
                             & $q_{23}$ & 0.000 & $q_{27}$ & 0.000 \\
                             & $q_{24}$ & 0.000 & $q_{28}$ & 0.000 \\
                                     
         \midrule
         \multirow{4}{*}{$\zeta_3$} & $q_{31}$ & 3.680 & $q_{35}$ & 2.329 \\
                             & $q_{32}$ & --3.905 & $q_{36}$ & --6.797 \\
                             & $q_{33}$ & 1.881 & $q_{37}$ & 1.000 \\
                             & $q_{34}$ & 2.000 & $q_{38}$ & --0.04118 \\
         \midrule
         \multirow{4}{*}{$\zeta_4$} & $q_{41}$ & 161.1 & $q_{45}$ & 1.034 \\
                             & $q_{42}$ & --7.690 & $q_{46}$ & --2.356 \\
                             & $q_{43}$ & 7.577 & $q_{47}$ & --0.001288 \\
                             & $q_{44}$ & 1.031 & $q_{48}$ & 1.000 \\
         \midrule
         \multirow{1}{*}{$\zeta_5$} & -- & --161.0 & -- & -- \\
         
         \midrule
         \multirow{1}{*}{$\zeta_6$} & -- & 8.286e--08 & -- & -- \\
                                     
    \bottomrule
    \end{tabular}
\end{table}

\renewcommand{\thetable}{\thesection\arabic{table}}
\begin{table}
    \caption{Fitting parameters, $k_{ij}$, used in Eqn.~\ref{eq:xi_i} to predict the near-field loss.}
    \label{table:NF_params}
    \centering
    \begin{tabular}{c|cccc}
    \toprule
    \toprule
         & Parameter & Value & Parameter & Value \\
         \midrule
         \multirow{4}{*}{$\xi_1$} & $k_{11}$ & --535.7 & $k_{15}$ & 0.0002700 \\
                             & $k_{12}$ & 843.4 & $k_{16}$ & --0.01872	 \\
                             & $k_{13}$ & --847.1 & $k_{17}$ & --0.3845 \\
                             & $k_{14}$ & 494.2 & -- & -- \\
                                     
         \midrule
         \multirow{4}{*}{$\xi_2$} & $k_{21}$ & 0.002909 & $k_{25}$ & 0.0002700 \\
                             & $k_{22}$ & 0.001348 & $k_{26}$ & 0.0000 \\
                             & $k_{23}$ & --0.001355 & $k_{27}$ & 0.0000 \\
                             & $k_{24}$ & --0.002975 & -- & -- \\
                                     
         \midrule
         \multirow{4}{*}{$\xi_3$} & $k_{31}$ & 4811 & $k_{35}$ & 0.0002700 \\
                             & $k_{32}$ & --0.03641 & $k_{36}$ & 0.0000 \\
                             & $k_{33}$ & --0.2382 & $k_{37}$ & 0.0000 \\
                             & $k_{34}$ & --4021 & -- & -- \\
                                     
    \bottomrule
    \end{tabular}
\end{table}

\begin{table}
    \caption{Fitting parameters, $s_{ij}$, used in Eqn.~\ref{eq:psi_i} to predict the far-field loss. Note, the $^*$ on $\psi_2^*$ indicates that this parameter is given in units of kg~MJ$^{-1}$ to ensure non-dimensionality of the expression.}
    \label{table:FF_params}
    \centering
    \begin{tabular}{c|cccc}
    \toprule
    \toprule
         & Parameter & Value & Parameter & Value \\
         \midrule
         \multirow{3}{*}{$\psi_1$} & $s_{11}$ & 1020 & $s_{14}$ & --1030 \\
                             & $s_{12}$ & 9.295 & $s_{15}$ & 0.00003300 \\
                             & $s_{13}$ & --0.1117 & $s_{16}$ & --0.05019 \\
                                     
         \midrule
         \multirow{3}{*}{$\psi_2^*$} & $s_{21}$ & --3.894 & $s_{24}$ & 4.993 \\
                             & $s_{22}$ & --1.102 & $s_{25}$ & 0.0005810 \\
                             & $s_{23}$ & --0.0004730 & $s_{26}$ & 0.003429 \\
                                     
         \midrule
         \multirow{3}{*}{$\psi_3$} & $s_{31}$ & --3208 & $s_{34}$ & 0.000 \\
                             & $s_{32}$ & 3209 & $s_{35}$ & 0.000 \\
                             & $s_{33}$ & 0.0003880 & $s_{36}$ & 0.000 \\
        \midrule
         \multirow{3}{*}{$\psi_4$} & $s_{41}$ & --322.5 & $s_{44}$ & 322.9 \\
                             & $s_{42}$ & --1.923 & $s_{45}$ & 0.0001800 \\
                             & $s_{43}$ & 1.415 & $s_{46}$ & --0.1320 \\
                                     
    \bottomrule
    \end{tabular}
\end{table}
\clearpage

\section{Moon-forming scenarios}\label{sec:MFI_scenarios}
\setcounter{table}{0}

\renewcommand{\thetable}{\thesection\arabic{table}}
\begin{table}[h]
\caption{Target and impactor masses ($M_{\rm t}^{\rm r}$, $M_{\rm i}^{\rm r}$), impactor-to-total system mass ratios (\textgamma), impact parameters ($b$), and impact velocities ($v_{\rm c}$) for the six Moon-forming scenarios considered in Section \ref{subsec:MFI}. Planet radii ($R_{\rm t}^{\rm r}$) were approximated as $M_{\rm t}^{{\rm r}^{1/4}}$.}
    \centering
    \begin{tabular}{l|ccccc}
    \toprule
    \toprule
        Model & $M_{\rm t}^{\rm r}$ (M$_{\oplus}$) & $M_{\rm i}^{\rm r}$ (M$_{\oplus}$) & \textgamma$\,(M_{\rm i}^{\rm r}/M_{\rm tot}^{\rm r})$ & $b$ & $v_{\rm c}$ $(v_{\rm esc})$ \\
        \midrule
         \citet{canup_origin_2001} (CA01) & 0.9070 & 0.1121 & 0.11 & 0.7 & 1.00 \\
         \midrule
         \citet{reufer_hit-and-run_2012} (R12) & 0.9000 & 0.2000 & $0.\overline{18}$ & 0.57 & 1.20 \\
         \midrule
         \citet{cuk_making_2012} (CS12) & 1.050 & 0.05000 & $0.0\overline{45}$ & 0.3 & 2.00 \\
         \midrule
         \citet{canup_forming_2012} (C12) & 0.5709 & 0.4671 & 0.45 & 0.55 & 1.10 \\
         \midrule
         \citet{lock2018origin} (LS18a) & 0.5720 & 0.4680 & 0.45 & 0.4 & 1.34 \\
         \midrule
         \citet{lock2018origin} (LS18b) & 0.7500 & 0.3000 & 0.2857 & 0.6 & 1.25 \\
         \bottomrule
    \end{tabular}
    \label{tab:MFI_scenarios}
\end{table}

\clearpage
\bibliography{references}{}
\bibliographystyle{aasjournalv7}

\end{document}